\documentclass[twocolumn,twocolappendix]{aastex631}

\usepackage{amsmath}
\usepackage{multirow}
\usepackage{subfigure}

\newcommand{\msun}{\ensuremath{\,\rm{M}_\odot}}

\newcommand{\ergs}{\ensuremath{\rm{erg/s}}}

\newcommand{\mint}{\ensuremath{\rm{M_{init}}}}

\newcommand{\mhe}{\ensuremath{\rm{M_{He}}}}

\newcommand{\crate}{\ensuremath{^{12}\rm{C}\left(\alpha,\gamma\right)^{16}\!\rm{O}}}

\newcommand{\tlookback}{\ensuremath{\log_{10}\; \left(\rm{\tau_{cc}-\tau}\right)/\rm{yr}}}

\newcommand{\teff}{\ensuremath{\rm{T_{eff}}}}

\newcommand{\code}[1]{\texttt{#1}}

\newcommand{\MESA}{\code{MESA}}

\newcommand{\kms}{{\mathrm{km\ s^{-1}}}}
\newcommand{\Msun}{{\mathrm{M}_\odot}}

\newcommand{\nuclei}[2]{\ensuremath{\mathrm{^{#1}#2}}}

\newcommand{\helium}[1][4]{\nuclei{#1}{He}}

\newcommand{\carbon}[1][12]{\nuclei{#1}{C}}

\newcommand{\oxygen}[1][16]{\nuclei{#1}{O}}

\newcommand{\neon}[1][20]{\nuclei{#1}{Ne}}

\newcommand{\zinc}[1][64]{\nuclei{#1}{Zn}}
\newcommand{\nickel}[1][56]{\nuclei{#1}{Ni}}

\begin{document}

\title{The cosmic carbon footprint of massive stars stripped in binary systems}

\author[0000-0003-3441-7624]{R.~Farmer}
\email{r.j.farmer@uva.nl}
\affiliation{Max-Planck-Institut für Astrophysik, Karl-Schwarzschild-Straße 1, 85741 Garching, Germany} 
\affiliation{Anton Pannekoek Institute for Astronomy and GRAPPA, University of Amsterdam, 
NL-1090 GE Amsterdam, The Netherlands}

\author[0000-0003-1009-5691]{E.~Laplace}
\affiliation{Anton Pannekoek Institute for Astronomy and GRAPPA, University of Amsterdam, 
NL-1090 GE Amsterdam, The Netherlands}

\author[0000-0001-9336-2825]{S.~E.~de~Mink}
\affiliation{Max-Planck-Institut für Astrophysik, Karl-Schwarzschild-Straße 1, 85741 Garching, Germany}
\affiliation{Anton Pannekoek Institute for Astronomy and GRAPPA, University of Amsterdam, 
NL-1090 GE Amsterdam, The Netherlands}
\affiliation{Center for Astrophysics | Harvard \& Smithsonian, 60 Garden Street, Cambridge, 
MA 02138, USA}

\author[0000-0001-7969-1569]{S.~Justham}
\affiliation{School of Astronomy \& Space Science, University of the Chinese Academy of Sciences, 
Beijing 100012, China}
\affiliation{National Astronomical Observatories, Chinese Academy of Sciences, Beijing 100012, China}
\affiliation{Anton Pannekoek Institute for Astronomy and GRAPPA, University of Amsterdam, NL-1090 
GE Amsterdam, The Netherlands}
\affiliation{Max-Planck-Institut für Astrophysik, Karl-Schwarzschild-Straße 1, 85741 Garching, Germany} 

\date{\today}

\begin{abstract}

The cosmic origin of carbon, a fundamental building block of life, is still uncertain.  Yield predictions for massive stars are almost exclusively based on single star models, even though a large fraction interact with a binary companion. Using the \MESA{} stellar evolution code, we predict the carbon ejected in the winds and supernovae of single and binary-stripped stars at solar metallicity. We find that binary-stripped stars are twice as efficient at producing carbon (1.5--2.6 times, depending on choices on the slope of the initial mass function and black hole formation).  We confirm that this is because the convective helium core recedes in stars that have lost their hydrogen envelope, as noted previously.  The shrinking of the core disconnects the outermost carbon-rich layers created during the early phase of helium burning from the more central burning regions. The same effect prevents carbon destruction, even when the supernova shock wave passes.  The yields are sensitive to the treatment of mixing at convective boundaries, specifically during carbon-shell burning (variations up to 40\%) and improving upon this should be a central priority for more reliable yield predictions. The yields are robust (variations less than 0.5\%) across our range of explosion assumptions. Black hole formation assumptions are also important, implying that the stellar graveyard now explored by gravitational-wave detections may yield clues to better understand the cosmic carbon production. Our findings also highlight the importance of accounting for binary-stripped stars in chemical yield predictions and motivates further studies of other products of binary interactions. 
\end{abstract}

%
\section{Introduction}\label{sec:intro}  

Understanding the Cosmic production of the elements that form the building blocks of life is still one of the main quests for modern astronomy. Massive stars are known to play a critical role in the synthesis of heavy elements over cosmic time, but many questions remain open \citep[e.g.][]{Burbidge57,cameron59,Woosley95, Nomoto06}.
Our current understanding of the nucleosynthesis products of massive stars is still almost exclusively  based on single star progenitor models \citep{maeder92,woosley93b,kobayashi06,curtis19}. However, observational studies of young massive stars have indicated that massive stars are nearly always born in multiple systems \citep{Abt1983, Mason+2009}, with at least one companion star nearby enough for binary interaction \citep{sana:12, Moe+2017}. Such interactions can dramatically change the final fate of massive stars \citep{podsiadlowski92,wellstein99,langer12}. How such interactions may change the chemical yields of massive stars is still not fully clear (see, however, \citealt{dedonder04,izzard04,izzard06}; \citealt{woosley19}). Models of $^{26}\rm{Al}$ nucleosynthesis find that binary stars can produce significantly different yields from single stars  \citep{braun95,brinkman19}.

The source of carbon in the Universe is still uncertain \citep{bensby06,romano17}. Observations, and theoretical modelling, suggest
contributions from the winds of asymptotic giant branch (AGB) stars \citep{nissen14}, massive star winds and their core 
collapse supernovae \citep{franchini20},
and type Ia supernovae \citep{leung20}. The source of carbon matters not just for the amount 
of carbon
expected in the Universe, but also for understanding when and where it is formed
\citep{carigi05,cescutti09}. Which can then be used to understand the star formation history of a
galaxy \citep{carilli13,romano20}.

Massive stars
are able to eject carbon a few million years after formation \citep{woosley93}, while AGB winds
and type Ia supernovae require
much longer timespans before releasing their carbon \citep{henry00,akerman04}. The relative
contribution from each source may vary over time as the metallicity, and thus wind
mass loss, increases \citep{dray03b,lau20}.

Carbon plays a crucial role in the interstellar medium (ISM) through its complex chemistry and its
ability to form a wide range of carbon-rich molecules \citep{herbst09} and carbonaceous dust \citep{li01,weingartner01}. 
Atomic carbon plays
key roles in heating and cooling interstellar gas \citep{wolfire95} as well as in tracing the properties of the ISM \citep{wolfire03}. 
The presence of CO is an important observational tracer of molecular gas \citep{frerking82,solomon87}.
Dust formation from supernovae is also governed by the presence of carbon
\citep{bevan17,sarangi18,lau20, brooker21}.
Thus understanding the formation of carbon and its distribution
is key to understanding the ISM \citep{burton78,gullberg18}.

Here we study the effect of binary evolution on the carbon yields ejected by massive stars \citep{langer91}.
Previous work by \citet{laplace20,laplace21} explored the evolution
of binary-stripped stars up to core collapse, and showed how the mass
loss during a binaries evolution alters the final structure of a star.
These structural differences in binary-stripped stars, as compared to single stars, 
is expected to lead to differences in the final
supernova and the yields \citep{woosley19,schneider21}.

We take this work further by exploring
the nucleosynthetic yield of carbon (\carbon{}) before and after core collapse and over a larger range of 
initial masses. We consider here the fate of binaries that are stripped by their companion in 
case B mass transfer, i.e they lose mass during their evolution after core hydrogen depletion
but before core helium ignition.
An analysis of all nucleosynthetic yields is deferred to later work.

Our paper is structured as follows,
in Section \ref{sec:meth} we describe our method for following the evolution 
of single and binary stars, as well as their supernova explosions. 
In Section \ref{sec:totalc12} we compute the carbon yields for single and binary stars. 
We discuss the uncertainties in our pre-supernova evolution and supernova explosions in Section \ref{sec:phys_var}.
In Section \ref{sec:imf}
discusses the initial mass function (IMF) weighted yields. Finally,
we discuss our results in Section \ref{sec:discus} and conclude in
Section \ref{sec:conc}.

\section{Method}\label{sec:meth}

\subsection{Pre-supernova evolution}

We use the MESA stellar evolution code
\citep[version 12115,][]{paxton:11,paxton:13,paxton:15,paxton:18,paxton:19} to evolve massive
single and binary stars from the zero-age main sequence to
core collapse. Our single stars and the primary (initially most massive star) in the binary have initial 
masses between $\mint=11$ -- $45\msun{}$. For
binary stars we set the initial period to be
between 38\textendash300 days.
This period range ensures that all binary stars undergo case B mass transfer \citep{paczynski67b}.
We set the secondary star's mass such that the mass ratio $\rm{M_2/M_1}=0.8$.  All models
are computed with a initial solar metallicity of Z=0.0142 and are non-rotating.
Single stars and binary stars 
are evolved with initial Y=0.2684 \citep[$\rm{Y=2Z+0.24}$][]{pols95,tout96}.
Inlists with all input parameters and models are made available online at \url{https://doi.org/10.5281/zenodo.4545836}.

To evolve the systems we build upon the method in \citet{laplace20,laplace21}. 
 We follow in detail
the structure of the primary star and the period evolution of the system during
Roche lobe overflow. 
We take the secondary star in the binary to be a point mass and do not follow its evolution. 
Mass transfer is assumed to occur 
conservatively such that no mass is lost from the system. With the initial periods chosen we do not expect
further Roche lobe overflow (RLOF) to occur during the systems lifetime \citep{laplace20}. 
After core helium burning ceases we evolve only the initial primary star of the binary. 

Wind driven mass loss follows the prescriptions of \citet{vink:01} for stars with $\teff>10^4K$ and surface hydrogen mass fraction $\rm{X_H}>0.4$, \citet{nugis:00} for 
$\teff>10^4K$ and surface $\rm{X_H}<0.4$, and \citet{dejager88} at all other times, with wind-scaling factors of 1.0. 
In binary systems we define all of the
mass lost when the radius of the primary is greater than its Roche lobe radius
to be RLOF, even though mass loss via winds will still occur. This is a reasonable
assumption as the mass loss via winds during Roche lobe overflow is small
due to the short timescale over which RLOF occurs.

Convective overshoot is calibrated to that of \citet{brott11}, with a step overshoot value of $f=0.385$ and $f_0=0.05$. In \MESA{} overshoot starts inside a convection zone at a distance of $f_0$ (in pressure scale heights),
and extends from this point a distance of $f$ (in pressure scale heights). Therefore, overshoot will extend a distance $f-f_0$ from the 
edge of the convective boundary.
We also apply the same amount of overshooting above the metal burning zones during the late stage evolution of the models.
We add a small amount of overshoot ($f=0.05$, $f_0=0.01$) below metal burning shells to improve numerical stability.
We use MLT++ for all models to improve the numerical stability of the low density envelopes \citep{paxton:13}. 
We include semiconvection with a mixing efficiency of $\alpha_{\rm{semi}}=1.0$ and we do not include thermohaline mixing.
Additional physics choices are specified in Appendix \ref{sec:other_phys}.

We evolve our stars with \MESA's \texttt{approx21.net} which contains
21 isotopes, following the alpha-chain up to iron. \citet{farmer16}
has shown the need to use larger nuclear networks when evolving stars to core-collapse, to compute the core structure accurately. 
However, we show in Appendix \ref{sec:comp} that models computed with \texttt{approx21.net} predict similar \carbon{} yields to models using such a larger network (\texttt{mesa\_128.net}).

We define the helium core mass of the star as the first point in time (and space) when the helium mass fraction, $
\rm{X_{He}}>0.1$ and the hydrogen mass fraction (at the same mass coordinate) is $\rm{X_{H}}<0.01$.
Core oxygen depletion is defined when the oxygen mass fraction at the center of the star drops below $\rm{X_{O}}<10^{-4}$.
Finally, we define core collapse to occur when the inner regions of the star infalls at $300\kms$.

\subsection{Core-collapse supernovae}

To model the supernova explosion, its shock, and the resulting nucleosynthesis we
place a ``thermal bomb'' at the center of our model \citep{aufderheide91,sawada19}.
First we excise a portion of the star's core,
the material that will form a compact object, by placing the inner boundary of our model
at the point where the entropy per baryon $S=4$ \citep{brown13}. We then inject energy into the
base  of the material outside this boundary over a mass range of $0.01\msun$, over 0.0045 seconds.
We inject
sufficient energy to bring the total energy of the star (the sum of the kinetic plus thermal energy minus the gravitational binding energy) to $10^{51}\ergs$. These values specify our default model assumptions.

This injection of energy then generates a hydrodynamic shock which travels from the inner boundary of the star to the surface. As it passes through the star it shock heats
material and begins nuclear burning. This nucleosynthesis is computed with \MESA's \texttt{mesa\_128.net} which contains 128 isotopes up to \zinc[64].
In Section \ref{sec:exp_phys} we discuss how our choice of explosion parameters
affects the resulting nucleosynthesis. Also in Section \ref{sec:exp_phys} we discuss our choice of
temporal and spatial resolution during the explosion. The star is then evolved until the shock reaches a location $0.1\msun$ below the surface, by
which time the shock has cooled to the point of no further nucleosynthesis occurring, except for
beta decays. We do not add by hand any \nickel{} to the stars.

During the shock propagation through the star we track the energy change due to photo-disintegrations and nuclear burning. At shock breakout our models will have a different final energy as compared to the amount of energy we injected during the explosion. The total energy is 1.05--1.20$\times10^{51}$ \ergs{}
at shock breakout while the kinetic energy at shock breakout is between 0.5--1.3$\times10^{51}$ \ergs.

We define the yield of an isotope as \citep{karakas16}:

\begin{equation}
    \rm{Yield} = \sum_T \Delta M_T \times \left(X_j - X_{j,int}\right)
\end{equation}

Where $\Delta \rm{M_T}$ is the mass lost over the time interval $\rm{T}$, $\rm{X_j}$ is the surface mass fraction of isotope $\rm{j}$, and $\rm{X_{j,int}}$ is the initial mass fraction of isotope $\rm{j}$. With this definition negative yields will occur in cases of net destruction of an isotope. In this case, the mass fraction in the ejected material will be lower than in the initial composition of the ejected material.
We use the solar composition of \citet{grevesse:98} which sets $\rm{X_{C_{12},int}=0.00244}$.

\begin{figure*}[ht]
  \centering
  \includegraphics[width=0.75\linewidth]{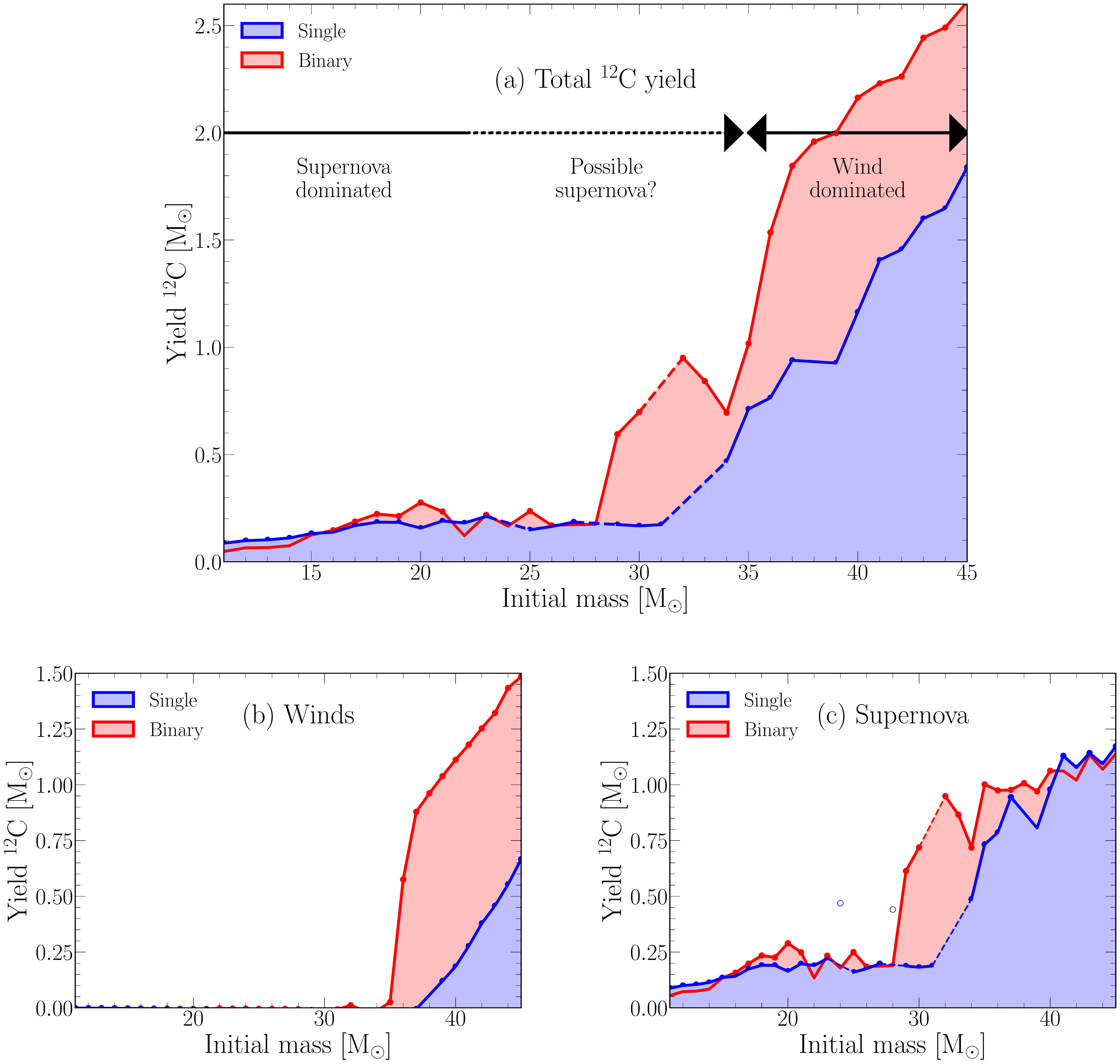}
  \caption{Top panel: The total \carbon{} yield from all ejection sources,
  Left Panel: The \carbon{} yield ejected during
  wind mass loss, Right Panel: The \carbon{} yield ejected during core collapse.
  The core-collapse yields assume all stars eject their envelope.
  In all panels red regions denote binary models while blue regions denote
  single star models.
  Open circles mark models which show anomalous carbon-burning behaviour, see section \ref{sec:cshell}. Dashed lines in panels (a) and (c) denote extrapolations over the
  anomalous carbon-burning behaviour and models which do not reach core collapse.
  The black arrows show the approximate location where each type of mass loss dominates the
  \carbon{} yield, taking into account reasonable assumptions for which stars eject their 
  envelopes \citep{sukhbold16,zapartas21b}.}
  \label{fig:yields}
\end{figure*}

\section{Total carbon}\label{sec:totalc12}

\begin{figure}[ht]
\includegraphics[width=\linewidth]{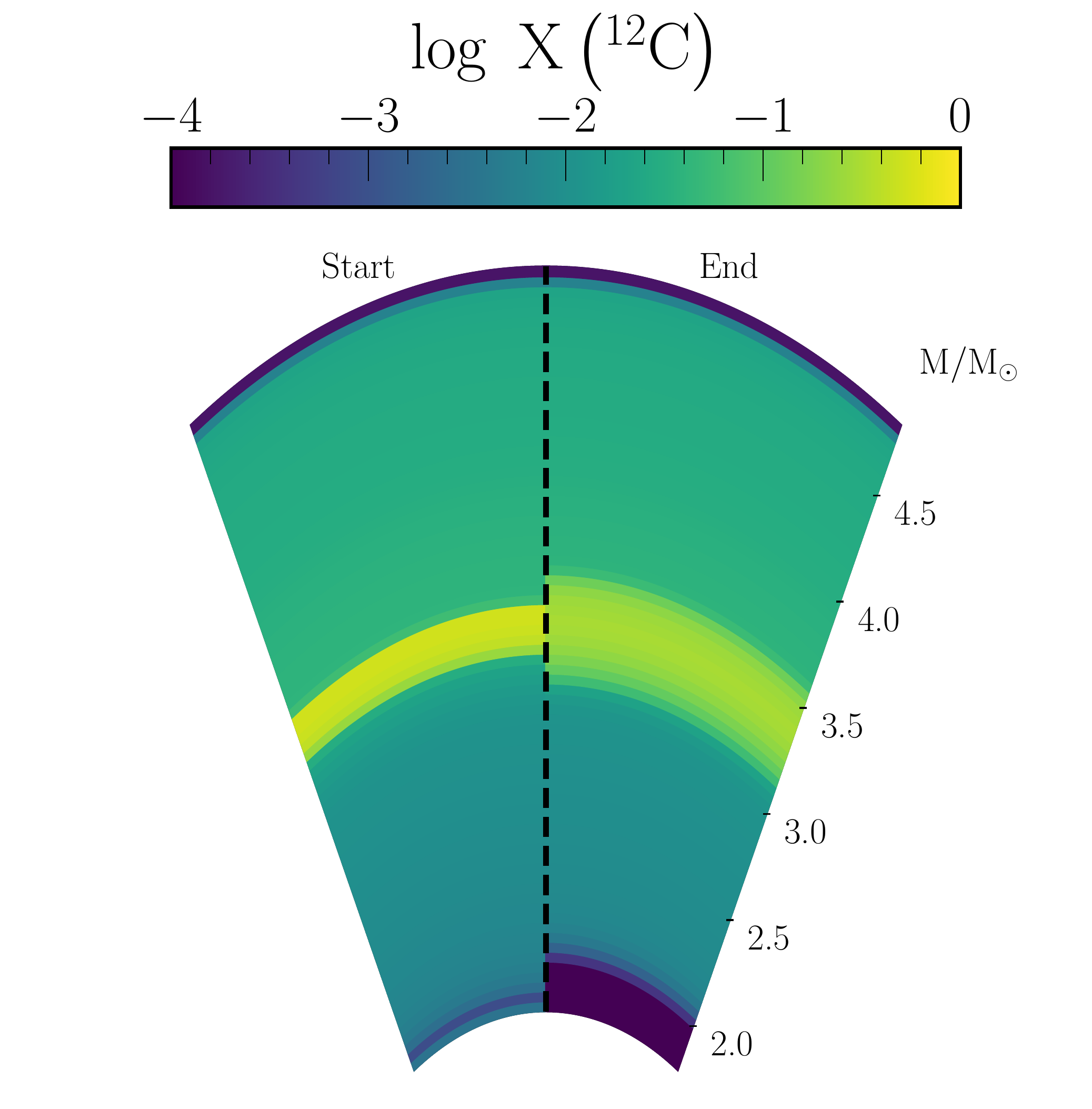}
\caption{The composition profile of a 15\msun{} binary-stripped star at the start and end of the shock propagation phase of
   core collapse. The colors denote the mass fraction of \carbon{}. 
   The animated version shows the shock propagation and resulting nucleosynthesis with time.} \label{fig:shock}
\end{figure}

Figure \ref{fig:yields}(a) shows the total \carbon{} yields from our single and binary stars, 
from all sources of mass loss.
Figure \ref{fig:yields}(b) shows the \carbon{} yield from winds, while Figure \ref{fig:yields}(c) shows the \carbon{} yield from
core-collapse ejecta. In appendix \ref{sec:table} we include table \ref{tab:yields} that breaks down the total carbon yield by its source.

The \carbon{} yield from the mass loss during RLOF is negative and small $\approx -0.01\msun$
and approximately independent of the initial mass of the primary star.
This is due to the envelope being unprocessed and its carbon content reflects this initial abundance.
The deeper layers of the envelope have been processed by CN and CNO-cycling and show depleted
carbon abundances \citep{maeder83}. The most massive stars show a slight decrease in \carbon{} ejected during RLOF as a larger fraction of the envelope is processed by CN(O)-cycling.

Figure \ref{fig:yields}(b) shows the \carbon{} yields due to wind mass loss.
The mass loss due to winds (for all stars) can be broken into two groups, for stars with $\mint\lessapprox35\msun$ their 
winds are not \carbon{} enriched as compared to their
initial composition, and thus not visible in Figure \ref{fig:yields}(b). Stars with $\mint\gtrapprox35\msun$
have \carbon{} enriched winds.
This is due to both the single and binary stars becoming fully stripped, removing both the hydrogen and helium layers of the star. 
In this initial mass range carbon rich material has been 
mixed out of the core and is then ejected. This occurs at $\mint=37\msun$ for single stars and $\mint=35\msun$ for the binaries. The transition occurs at a lower initial mass for the binaries due to RLOF removing some of the envelope.
Lower mass objects which are not fully 
stripped do not eject more \carbon{} in their stellar winds than they started with. 

Figure \ref{fig:yields}(c) shows the \carbon{} yields for core-collapse ejecta, assuming 
that all stars 
eject their envelopes. The \carbon{} yield is relatively flat as a function of initial mass for stars with 
$\mint \lessapprox 27\msun$, 
at $\approx0.2\msun$. Above this 
initial mass the yield rapidly increases up to $\approx1.25\msun$. 
This transition between low and high 
carbon yields occurs between $27\lessapprox \mint/\msun \lessapprox 35$.
The increased \carbon{} yields are
due to the wind mass loss removing the hydrogen envelope but not all of 
the helium envelope from the stars.
Thus there is enough mass loss to alter the core structure of the star, 
leaving behind a carbon layer
but not enough wind mass loss to then expose that layer (see Section \ref{sec:core}).

\subsection{Shock nucleosynthesis}

Figure \ref{fig:shock} shows the distribution of \carbon{} inside a binary-stripped star with $\mint=15\msun$
at the start and end of the shock propagation phase during core collapse. 
The online animation shows 
the propagation of the shock and the explosive nucleosynthesis this generates. 

At the start of the core-collapse phase, we can see that the \carbon{}
is concentrated in the helium shell, with only small amount of material near the core.
As the shock propagates it photo-disintegrates some of the \carbon{} near the inner boundary, and cools from a peak temperature of 
$\approx10^{10}\rm{K}$, to $\approx 3\times 10^{8}\rm{K}$ when it reaches the carbon-rich helium shell, 
and to $\approx10^{7}\rm{K}$ near shock breakout. 
By the time the shock reaches the carbon-rich 
layers it has cooled sufficiently that it can no longer burn those layers. The difference in
pre- and post- supernova nucleosynthesis is $\approx1\%$ for \carbon{}. The main effect on the \carbon{} yield
is to smear the \carbon{} distribution over a slightly larger range of mass coordinates. 
Thus for future studies it is 
not necessary to model the supernova explosion itself to predict \carbon{} yields. 
This was already found for single-star progenitors \citep[e.g][]{thielemann96,young07}, and we now confirm it for self-consistent binary-stripped structures.

\begin{figure*}[ht]
  \centering
  \includegraphics[width=1.0\linewidth]{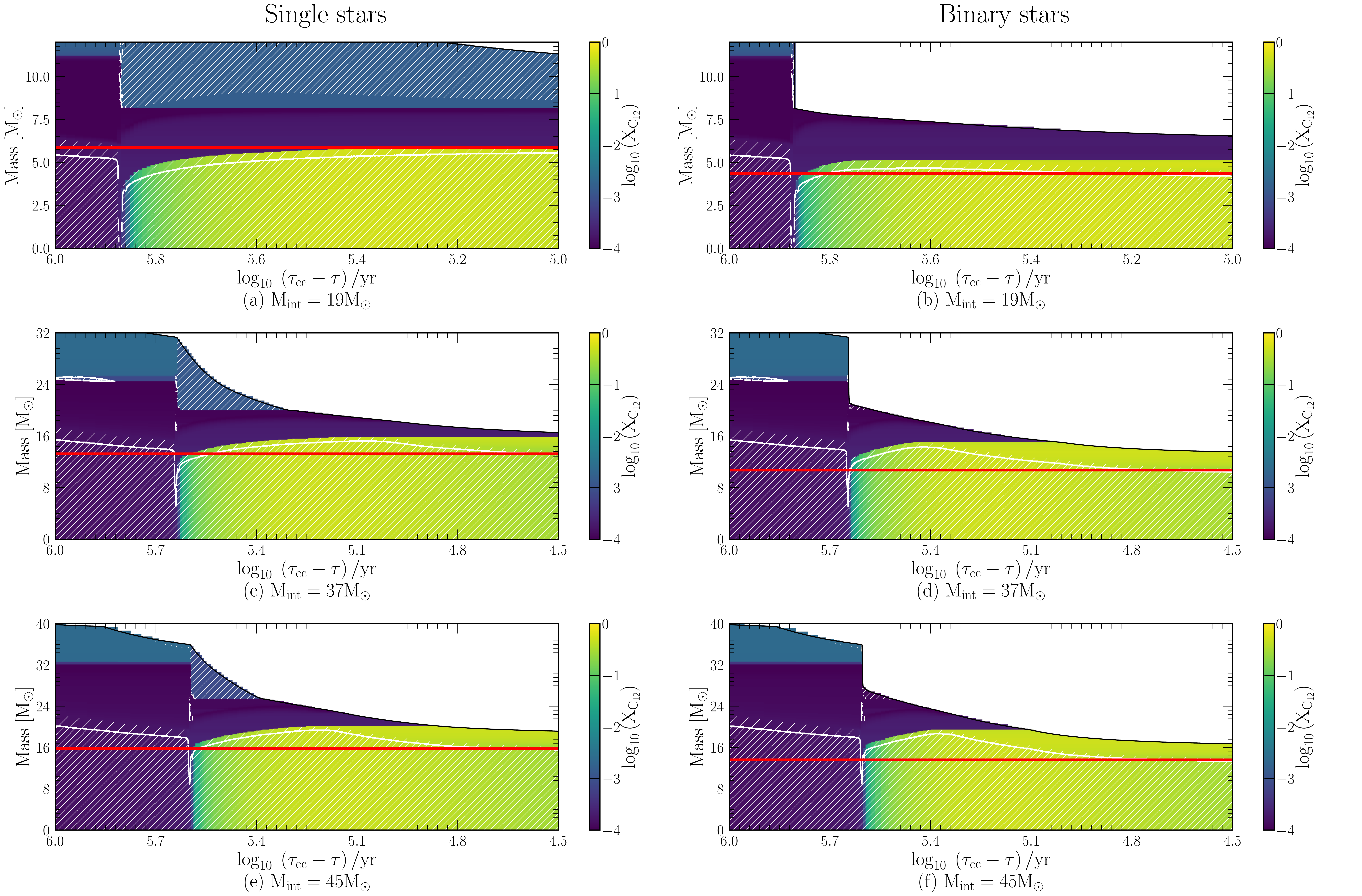}
  \caption{Kippenhahn diagrams for the inner regions of single and binary-stripped stars with 
  $\mint=19$, $37$, and $45\msun$, during core hydrogen and core helium burning. The left column shows single star models, while
  the right column shows binary-stripped models. The x-axis shows the
  time until core-collapse. Colors show the mass fraction of \carbon{} at each mass co-ordinate. Hatching shows mixing regions due to convection and overshoot.  The red horizontal line shows the mass coordinate for what will become the CO core
  at the end of core helium burning.}
  \label{fig:kipps}
\end{figure*}

\subsection{Differences in core structure}\label{sec:core}

To understand the differences between single stars and stripped binaries in Figure \ref{fig:yields} it is instructive to examine the evolution of
two stars (a single and a binary-stripped) of the same initial mass. Figure \ref{fig:kipps} shows the time evolution of a single and a binary-stripped star with $\mint=19,\ 37, \rm{and}\ 45\msun$. 
Considering the $\mint=19\msun$ case, both stars start on the main sequence and lose only a very small amount of mass before they evolve into Hertzsprung gap stars. At this point the donor star in the binary exceeds its Roche lobe radius and begins losing mass, via RLOF. In 
Figure \ref{fig:kipps}b this occurs at $\tlookback\approx 5.9$. Most of the binary star's envelope is lost at this point, bringing its mass down to $\approx 8.0\msun$, but the RLOF does not completely remove the hydrogen envelope \citep{gotberg17,yoon:17}.

However, the mass that was lost was only comprised of the stellar envelope (light blue region), which
has a composition similar to the star's birth composition. Thus the mass loss from the binary during RLOF does not enrich the Universe in \carbon{}. In fact, the material is slightly carbon
poor, due to some of the material having been CNO processed before the outer edge of the convective hydrogen-burning core receded (dark blue region).

As the binary loses mass from its outer layers the core structure of the binary is altered. The binary-stripped star forms a smaller helium core, and the edge of the convective core recedes during core helium burning \citep{langer89,woosley93}, while in the $\mint=19\msun$ single star
the mass in the convective helium-burning core stays constant. This receding convective core leaves
behind \carbon{} that was produced by the $3\alpha$ process but had not yet been converted into
oxygen in $\crate$ \citep{langer91}. This left-over \carbon{} is outside of what will become the CO core of the star (denoted by the red lines).

Any \carbon{} that is produced, and stays, inside the helium core will either be burnt
into oxygen at the end of core helium burning, destroyed during $\carbon{}+\carbon{}$ burning, or will be accreted onto the compact object during
core-collapse. Thus only the \carbon{} that is mixed out of the core has a chance to survive until core-collapse. 
Some of this outwardly mixed \carbon{} will not, however, survive until core collapse as it will be converted into oxygen during helium shell burning, or be mixed back into the core during carbon burning via the carbon convection zone intersecting the helium shell (see Section \ref{sec:cshell}).

Figure \ref{fig:kipps}(c,d) shows a single and binary-stripped star with $\mint=35\msun$.
Here the wind mass loss is sufficient to remove the remaining helium layers above the core. This
exposes \carbon{} rich material, which is then ejected in a wind, similar to the process in the 
binary-stripped star with $\mint=19\msun$.
Thus the single star has a 
net positive \carbon{} yield for its winds. The mass loss due to winds is also 
now strong enough in the single-star case to force the helium core to recede.

Figure \ref{fig:kipps}(e,f) shows stars
with $\mint=45\msun$. At this initial mass the winds of the single star are sufficient both to cause the helium core to recede
and to expose the \carbon{} rich layers to be ejected in a wind. Though this occurs at a slightly later time than in the binary star case
(as the binary-stripped star lost some mass in RLOF), thus there is less time for the star to eject this carbon rich material leading to 
a lower total carbon wind yield for the single star. 

\subsection{Carbon shell burning}\label{sec:cshell}

\begin{figure}[h]
  \centering
  \includegraphics[width=1.0\linewidth]{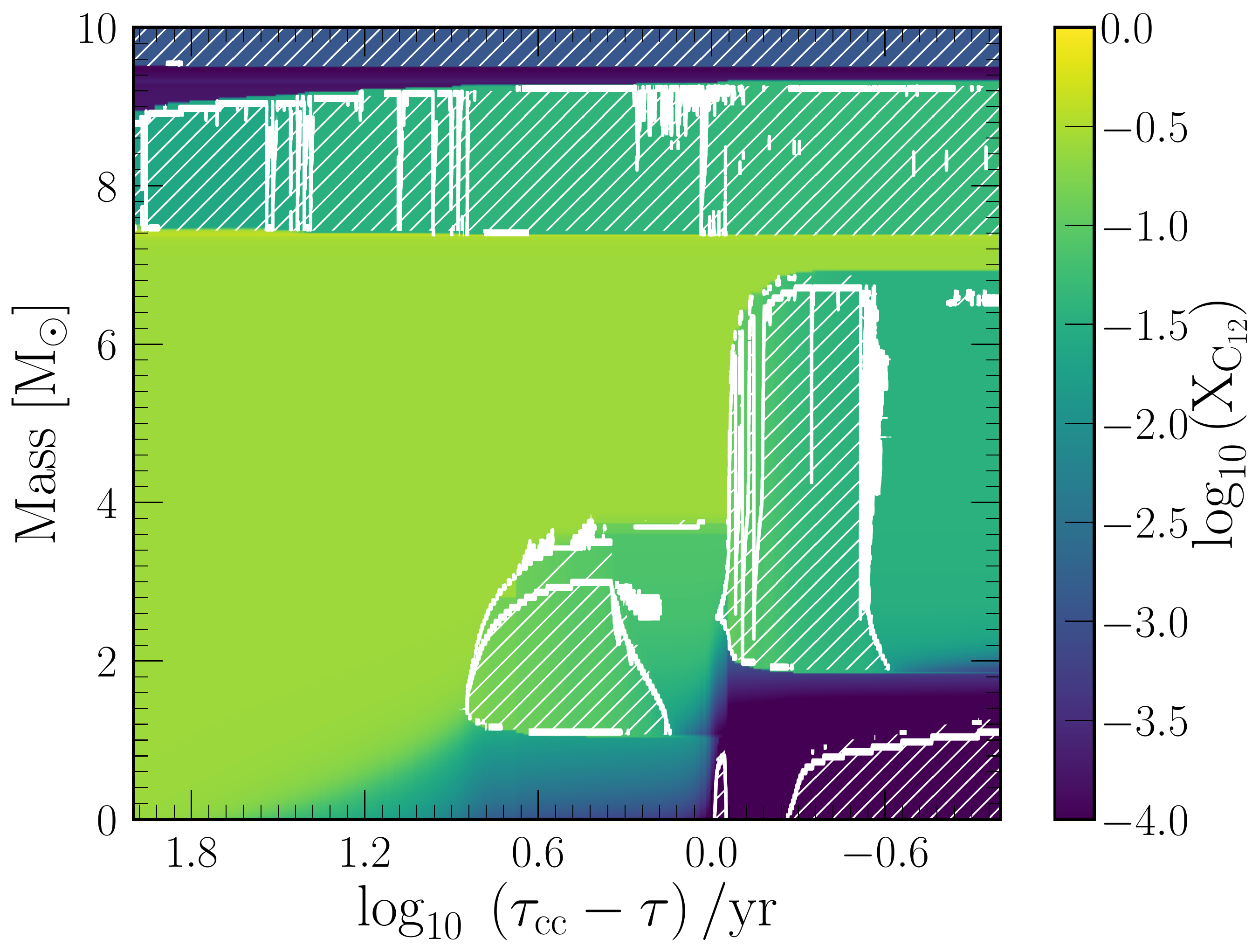}
  \caption{Kippenhahn diagram for a 23\msun{} single star during carbon and oxygen burning. Colors show the mass fraction of \carbon{} at each mass co-ordinate. Hatching shows mixing regions due to convection and overshoot. }
  \label{fig:kipp_c12}
\end{figure}

In Figure \ref{fig:yields} there is significant noise in the \carbon{} yields from core-collapse ejecta, especially at the higher masses. There are also several models marked with open symbols in 
the right panel of Figure \ref{fig:yields}. These are $\mint=24$, $28$, and $33\msun$ for the single stars and $\mint=31\msun$
for the binary-stripped stars. These variations occur due to changes in the behaviour of carbon burning shells during the 
star's carbon and oxygen burning phases. 

During the carbon shell burning phase, carbon initially ignites either at the center of the star or off-center and burns outwards \citep{arnett69,sukhbold20}. 
Figure \ref{fig:kipp_c12}
show the ignition of carbon in a 23\msun{} single star. This star ignites \carbon{} at the center at $\tlookback\approx 1.2$, radiatively,
this burning then moves outwards before it begins driving an off-center convection zone. This initial burning 
phase stops, before an additional \carbon{} burning zone ignites off-center, at $\tlookback \approx 0.0$ at the same time as the core
ignites \neon{}. It is this burning zone that causes the variability in the \carbon{} yields.

As this convection zone extends outwards, it mixes \carbon{} from the outer layers of the core inwards where the \carbon{} is then burnt. Thus the
maximal extent of this zone and the amount of time that it has to mix \carbon{} downwards sets the final \carbon{} yields.
Figure \ref{fig:kipp_c12} shows that in this model a small pocket of \carbon{} survives between the outer edge of the carbon convection zone and
the lower edge of the helium burning shell \citep{laplace21}. This pocket is a mix of \carbon{} and \helium{} (left over when the helium core receded at the end of core
helium burning). At core collapse, we find that for most models this pocket will contribute $\approx50\%$ of the final \carbon{} yield, with the remaining \carbon{} yield coming from \carbon{} produced in the helium shell. 

We expect that differences in the treatment of mixing boundaries during carbon-shell burning are also important for understanding the differences between the results obtained with different codes, see Appendix~\ref{sec:others}.

\section{Physics variations}\label{sec:phys_var}

\subsection{Sensitivity to physics choices}\label{sec:presn_sen}

The sensitivity to the size and timing of the carbon-burning shells may suggest that our models are under resolved, or that our choice
of convective overshoot above the carbon shell is significantly expanding the size of the convection zone. 
To test this
we ran two grids of additional models. Firstly we ran models for with a $23\msun$ single star model varying our
resolution controls. 
Next we randomly varied our choice of overshoot
controls ($f$ and $f_0$) above the carbon shell for the same $23\msun$ single star model (with our
default resolution controls).  We varied $f$ and 
$f_0$ between 0.0 and 0.05, with $f_0<f$.
For both sets of models we keep our default model assumptions until the end 
of core helium burning, where we then
change our model assumptions and evolve the models until core oxygen depletion.

We randomly picked values for each of the following spatial resolution controls, \texttt{mesh\_delta\_coeff} (0.5--1.0), \texttt{mesh\_delta\_coeff\_for\_highT} 
(0.5--1.0), and \texttt{max\_dq} ($10^{-4}$--$10^{-3}$). For our temporal controls 
we varied \texttt{dX\_nuc\_drop\_limit} ($10^{-4.5}$--$10^{-3.0}$) and
\texttt{varcontrol\_target} ($10^{-4.5}$--$10^{-3.0}$). With our choices we 
increase the spatial 
resolution by up to a factor of 5 and increase (decrease) the temporal resolution
by a factor of 3 (10).
This leads to the highest resolution models having $\approx 100,000$ time steps 
(for just carbon and oxygen burning) and 
$\approx 15,000$ mesh points.
Overall, we are primarily sensitive to the choice of \texttt{dX\_nuc\_drop\_limit},
which limits the 
timestep based on the rate of change of the most abundant isotopes, primarily 
variations in the \carbon{} and 
\oxygen{} abundances at the center of the star.

\begin{figure}[ht]
     \centering
     \subfigure[Temporal resolution]{\label{fig:coreotime}\includegraphics[width=1.0\linewidth]{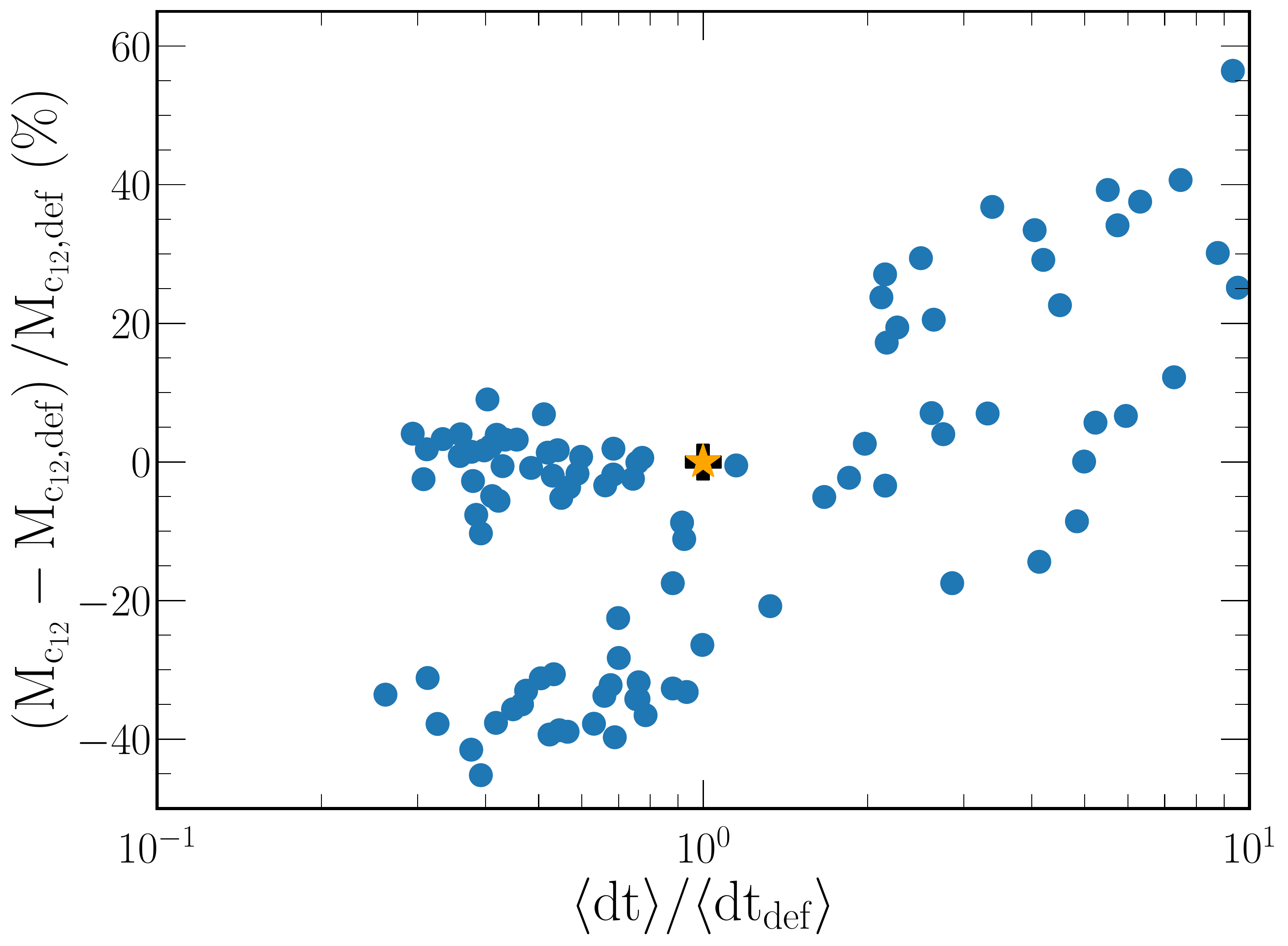}}
     
     \subfigure[Overshoot]{\label{figcoreoover}\includegraphics[width=1.0\linewidth]{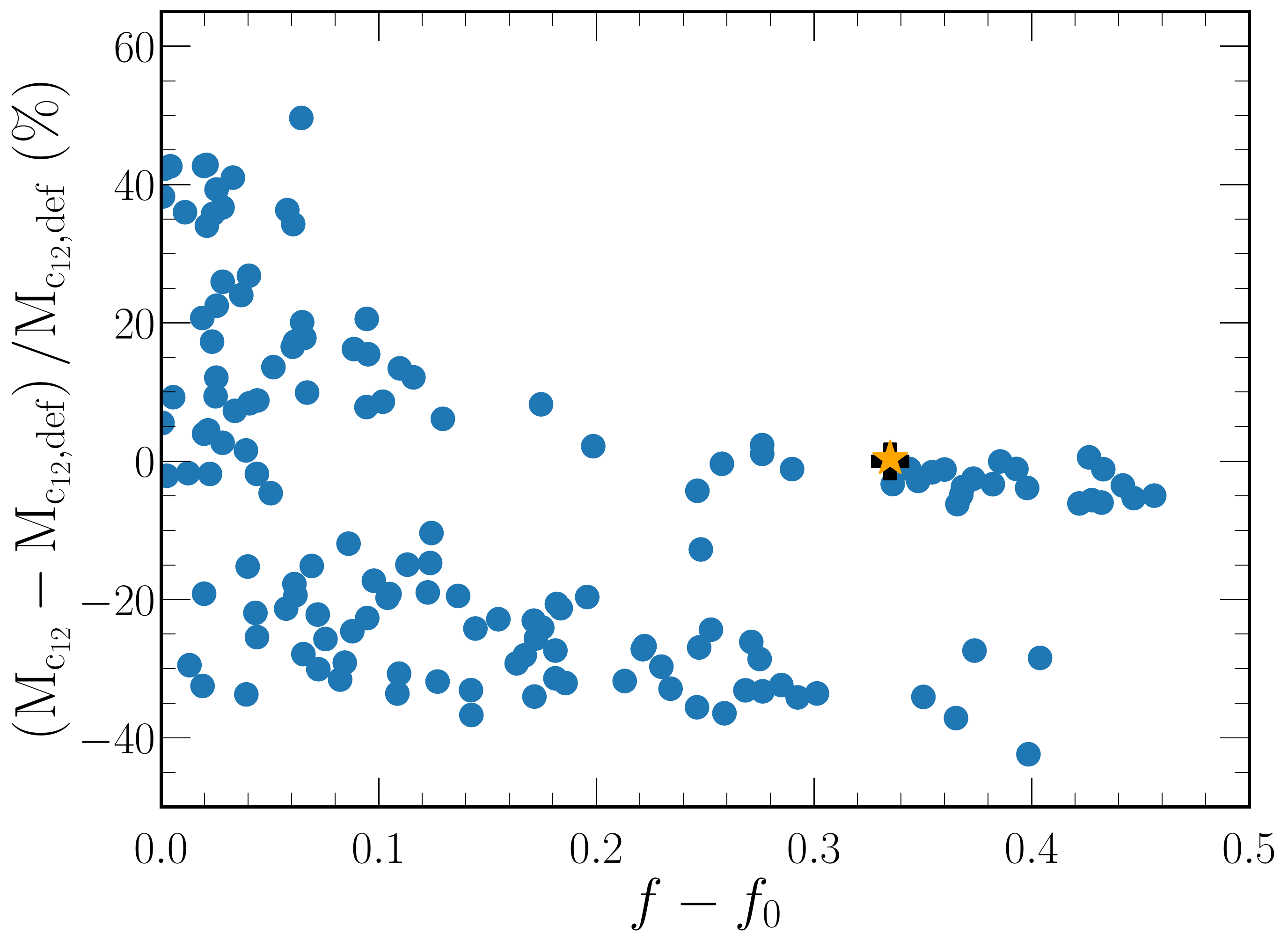}}
        \caption{The relative change in the final total mass of \carbon{} in the star measured at the point of core
        oxygen depletion, as a function of the temporal resolution and strength of convective overshoot for a $23\msun$ single star. 
        In panel (a) we vary a range of \MESA's temporal and spatial resolution controls. In panel (b)
        we vary both of the overshoot parameters $f$ and $f_0$. The x-axis in panel (a) is the average timestep relative to our default model, where the resolution increases towards the left.
        The x-axis in panel (b) is the physical extent of convective overshoot beyond the convective boundary above carbon burning shells. 
        In both panels the orange star denotes our default model assumptions, and the black plus symbol denotes
        a model which was evolved with \texttt{mesa\_128.net} from the ZAMS (with otherwise default assumptions).}
        \label{fig:coreo_c12}
\end{figure}

Figure \ref{fig:coreo_c12} shows the relative change in the total mass of \carbon{} in our model measured at the end of the 
core oxygen depletion for both variations in the resolution and the amount of convective overshoot above the carbon burning shell.
The black plus symbols in Figure \ref{fig:coreo_c12} show a comparison with a model ran with \texttt{mesa\_128.net}
from the ZAMS, which has only a $\approx 2\%$ difference in the total mass of \carbon{} (measured at core oxygen depletion).

Figure \ref{fig:coreotime} shows the relative change in the total mass of \carbon{} compared to
the relative change in the average time step taken. As the time resolution increases the total mass of \carbon{} at the end of 
core oxygen burning decreases,
relative to our default choices. However, even with at our highest resolutions, there is still a large spread possible in the 
final mass of \carbon{} ($\approx40\%$). We do not plot the spatial resolution variations as they
show no correlation with the carbon yields. 

Figure \ref{figcoreoover} shows the relative change in the \carbon{} yields
as a function of the amount of overshoot beyond the top of the carbon burning shells convective boundary (with all other 
overshoot regions keeping their same values). 
There is a slight trend for both the final total mass of \carbon{} to decrease, and the 
spread in \carbon{} values to decrease, as the amount of overshoot increases. 
However, it is possible to achieve the same final total mass of \carbon{} as our default model with the full range of overshoot values considered here.
While this shows we may not be sensitive to the overall amount of overshoot (above the carbon burning shells), there is considerable scatter in the final total mass of \carbon{} with the 
the same choice of the amount of overshoot. Thus \MESA{} users should consider carefully their
physical choices for the amount of overshoot ($f-f_0$) and the individual numerical choices ($f$ and $f_0$) needed to achieve this value. 

More work is needed to understand convection during 
this burning phase \citep{cristini17}, as well as the role of convective overshoot that can enhance this effect by extending the region over 
which the carbon convection zone can mix. We may be seeing a similar effect as shown in 
\citet{paxton:18} and \citet{paxton:19} with the improper placement of the convective boundary. 
We tested with both the predictive mixing \citep{paxton:18}
and convective pre-mixing schemes \citep{paxton:19}. Both schemes still show scatter in how large the carbon convective shell grows,
and for how long it is able to mix \carbon{} into the core.
We note however that both predictive mixing and convective pre-mixing
assume that the mixing timescale is shorter than the computational timestep, 
which breaks down during carbon burning.

\subsection{Sensitivity to the explosion properties}\label{sec:exp_phys}

\begin{figure*}[ht]
\centering

  \subfigure[Injected energy]{\label{fig:mc_eng}\includegraphics[width=0.45\linewidth]{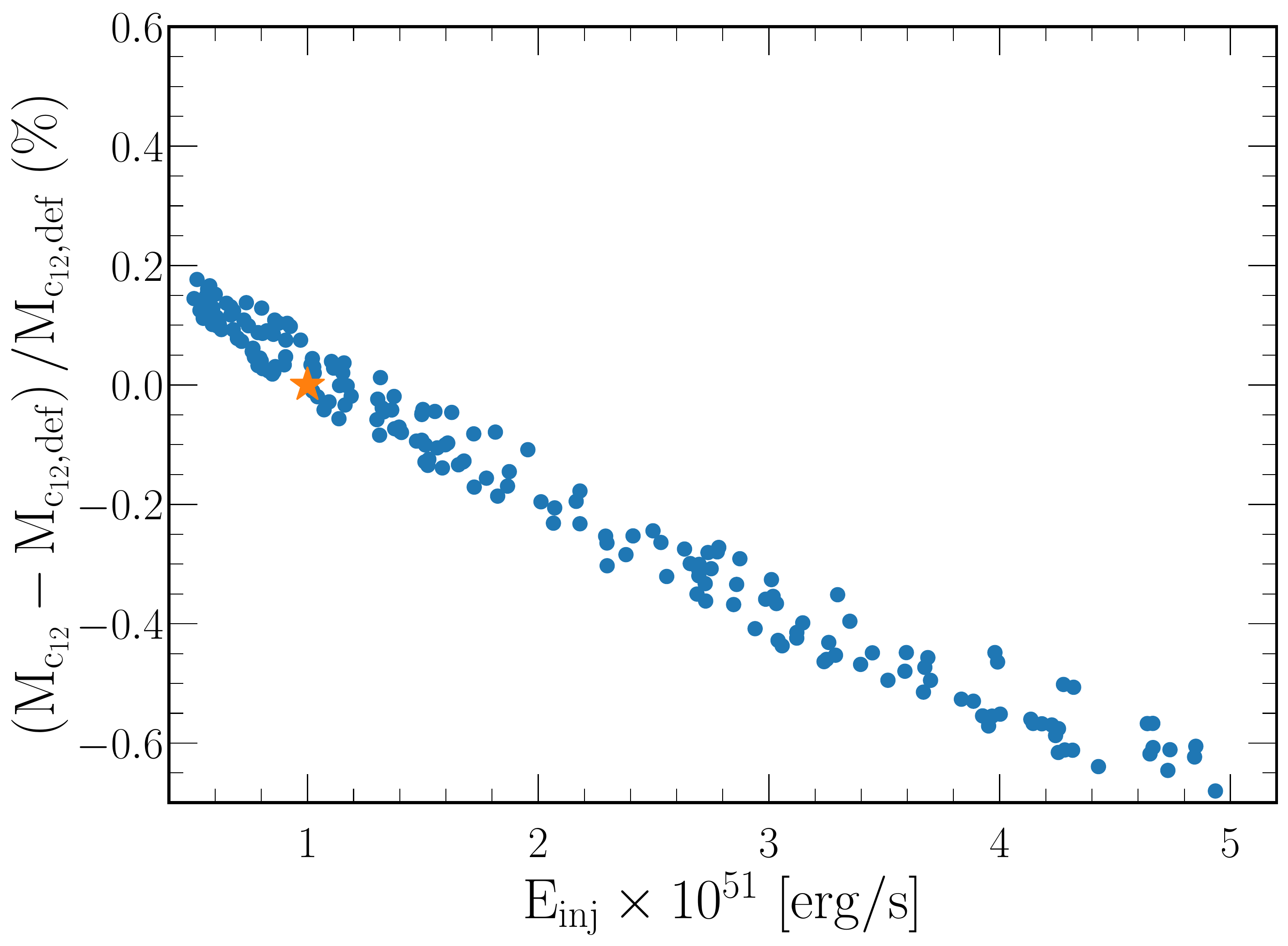}}
  \subfigure[Mass cut]{\label{fig:mc_mc}\includegraphics[width=0.45\linewidth]{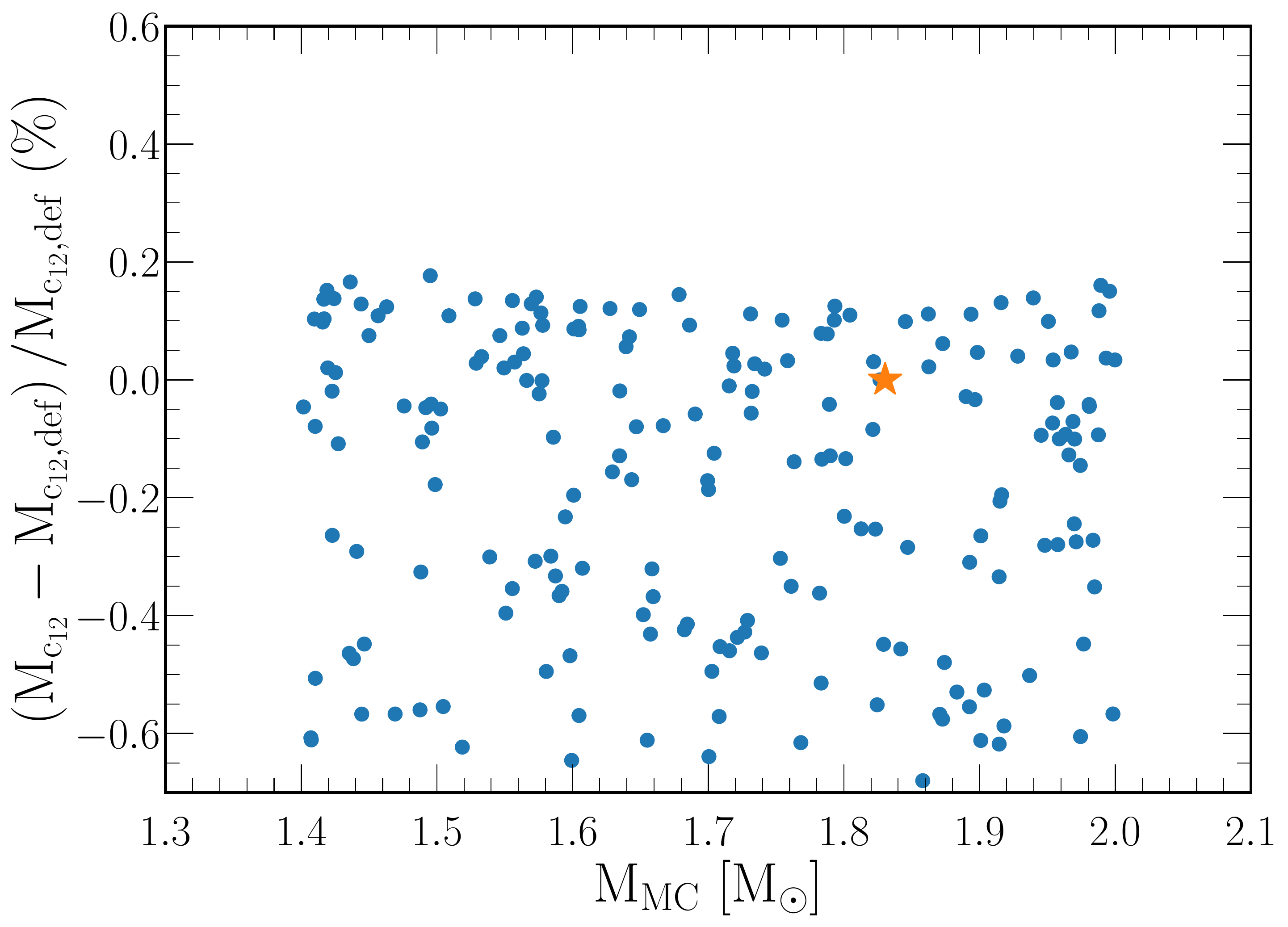}}
  
  \subfigure[Injection time]{\label{fig:mc_injtime}\includegraphics[width=0.45\linewidth]{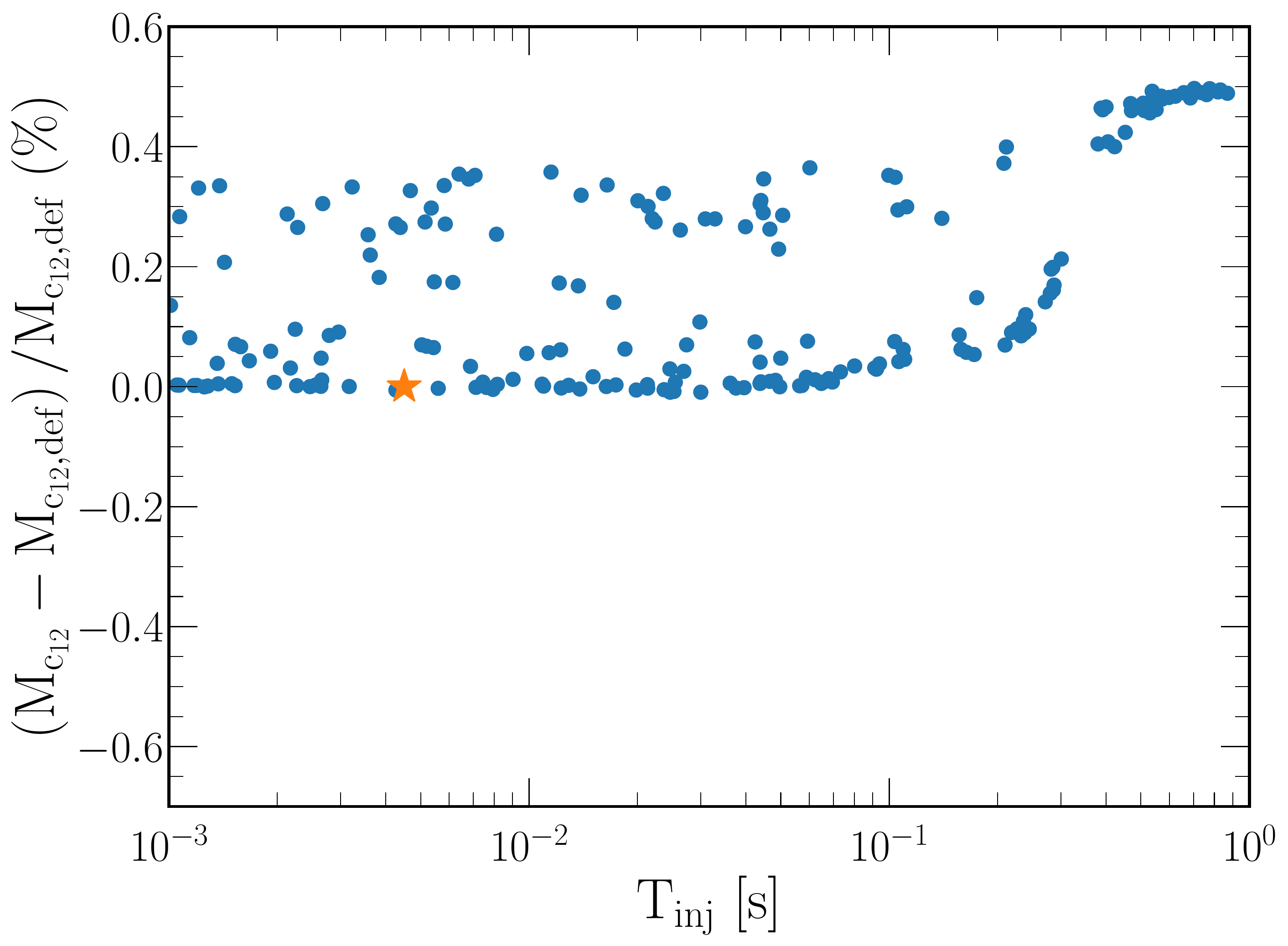}}
  \subfigure[Injection mass]{\label{fig:mc_injmass}\includegraphics[width=0.45\linewidth]{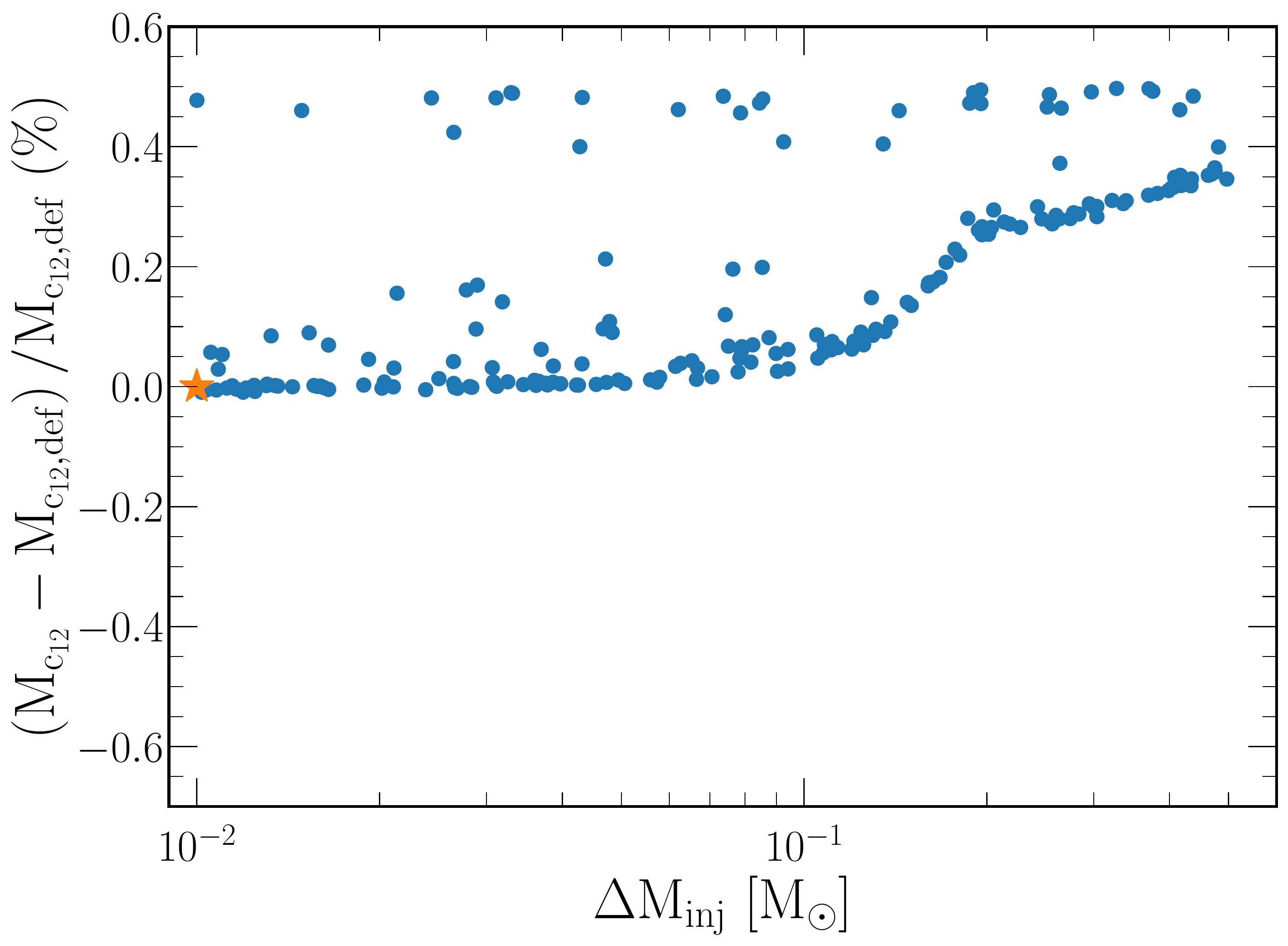}}  
  
  \subfigure[Spatial resolution]{\label{fig:mc_space}\includegraphics[width=0.45\linewidth]{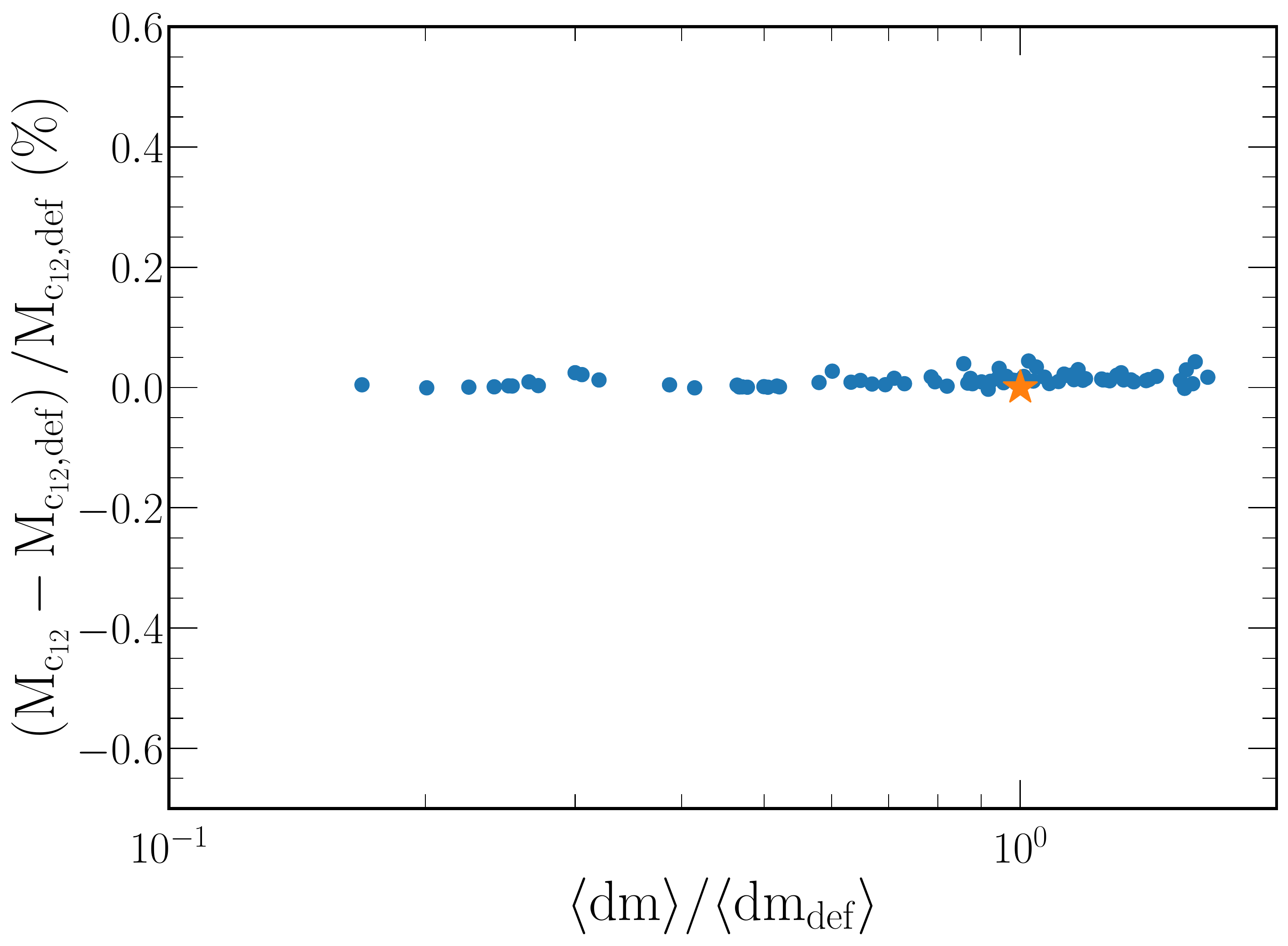}}
  \subfigure[Temporal resolution]{\label{fig:mc_time}\includegraphics[width=0.45\linewidth]{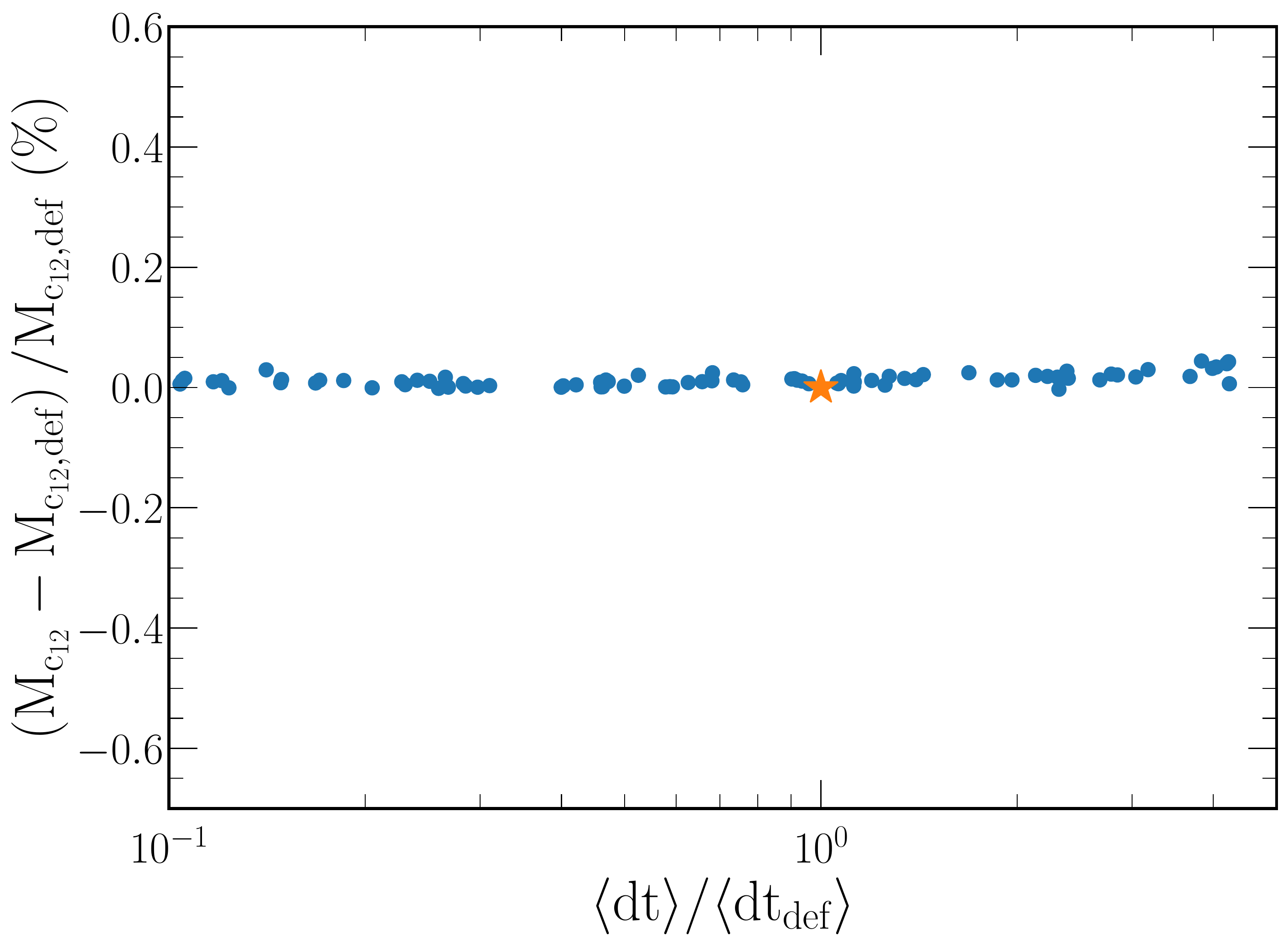}}

  \caption{The relative difference in the amount of \carbon{} ejected by supernovae of our binary-stripped 
  star model with $\mint=16\msun$, compared to our default ($\rm{def}$) model. 
  Panel (a) shows the variation with respect to the injected energy, 
  Panel (b) shows the variation with respect to the mass cut, 
  Panel (c) shows variations with respect to the time the energy is injected over,
  Panel (d) shows variations with respect to the range of masses that the energy is injected over,
  Panel (e) shows the variation with respect to changes in the spatial resolution, 
  and Panel (f) shows the variation with respect to changes in the temporal resolution.
  $\langle dm \rangle$ is the time-weighted average cell mass in $\msun$, $\langle dt \rangle$ is the average timestep in seconds. The spatial and temporal resolution increases to the left. 
  For each row, we simultaneously Monte Carlo sampled both the parameters shown  while keeping
  the other explosion parameters set to our  default assumptions.
The orange star denotes our default assumptions.}
  \label{fig:mc_sn}
\end{figure*}

To explore the sensitivity of our results to the assumptions made we perform two sets of tests on 
a $\mint=16\msun$ binary-stripped star. We do not extensively test the pre-supernova physics variations as they have previously been explored in \citet{farmer16,renzo:17,laplace21}.
Figure \ref{fig:mc_sn} shows the effect of varying the 
physics assumptions made during the core-collapse supernovae and the effects of varying the numerical assumptions made during the core-collapse supernova. Each row of Figure \ref{fig:mc_sn}
tests two physics/numerical assumptions at a time. The vertical spread in the distribution of points at a fixed x-coordinate shows 
how sensitive the \carbon{} yield is to the \emph{other} parameter shown in the same row. Thus a tight correlation indicates that the \emph{other} parameter in the same row only minimally affects the predicted \carbon{} yield. A horizontal line implies that the \carbon{} yield is insensitive to \emph{that} parameter. Negative 
$\rm{\left(M_{c_{12}}-M_{c_{12},def}\right)/M_{c_{12},def}}$ values indicate that 
more \carbon{} was destroyed than in our default model, while positive values indicate that 
less \carbon{} was destroyed.

In Figure \ref{fig:mc_eng} and \ref{fig:mc_mc}  we randomly sampled $\approx200$ times both the injected energy (between $0.5$ -- $ 5\times 10^{51} \ergs$) and the mass cut (between $1.4$ -- $ 2.0\msun$). Figure \ref{fig:mc_eng} shows the effect of varying the injected energy on the total mass of \carbon{} ejected during the supernovae. We can see a strong correlation between
injected energy and \carbon{} ejected, increasing the energy decreases the \carbon{} mass ejected. However, the change is $<1\%$ in the total mass of \carbon{} and in the context of Figure \ref{fig:yields} the change would only be of the order of the size of the symbols.

Figure \ref{fig:mc_mc} shows the effect as we vary the mass cut. We find no correlation with the mass of ejected \carbon{} for this range of mass cuts. These trends can be explained as follows: most of the \carbon{} resides in the helium shell and only a small amount has been mixed down 
(or produced during late stage burning) to near where the compact object will form. Thus changing the mass cut has little
effect, as the total mass of \carbon{} ejected is much greater than the change possible by moving the
inner boundary. 

The injected energy has a correlated (but small) effect on the \carbon{} yield due to the shock
processing the small amount of \carbon{} that exists near the inner boundary (see Figure \ref{fig:shock}). 
However by the time the shock reaches the bulk of the \carbon{} in the helium shell the shock has cooled below $10^8$K, thus
it can no longer burn the \carbon{} present.

In Figure \ref{fig:mc_injtime} and \ref{fig:mc_injmass}  we randomly sampled $\approx200$ 
times both the 
time over which we inject energy during the explosion ($10^{-3}<\rm{T_{inj}}/s<10^{0}$) and the mass over 
which we inject energy ($10^{-2}<\rm{\Delta M_{inj}}/\msun<5\times 10^{-1}$).
When either the $\rm{T_{inj}}$ is small ($\rm{T_{inj}}<0.1$s) or $\rm{\Delta M_{inj}}$ is small ($\rm{M_{inj}}<0.1\msun$) their effects on the \carbon{} yield is small. Only once $\rm{M_{inj}}$ exceeds
$0.1\msun$ does it begin to dominate the \carbon{} yield. However, once
$\rm{T_{inj}}>0.3$s it begins to dominate the \carbon{} yield instead.
When either $\rm{T_{inj}}$ or $\rm{\Delta M_{inj}}$ increases the 
generated shock is weaker as either its power is lower (as it spreads over more time) or its
deposited energy per unit mass is lower (as its spread over more mass) . This weakening of the
shock lowers the amount of
\carbon{} destroyed in the initial burning phases of the explosion \citep{sawada19}. The variations here are 
comparable to the uncertainty in the injected energy (though with an opposite sign).

For Figure \ref{fig:mc_space} and \ref{fig:mc_time}  we randomly sampled $\approx100$ times
both the temporal controls and the mesh 
controls, simultaneously, to probe the sensitivity of our predictions to the numerical resolution of our models. 
We varied a series of mesh controls (\texttt{split\_merge\_amr\_nz\_baseline} (500--8000),
\texttt{split\_merge\_amr\_MaxLong} (1.1--2.5), and  \texttt{split\_merge\_amr\_MaxShort} 
(1.1--2.5)), during the explosion, as seen in Figure \ref{fig:mc_space}. These 
controls force \MESA{} to distribute its mesh 
according to the radius of each zone (See section 4 of \citet{paxton:18}). We varied 
\texttt{dt\_div\_min\_dr\_div\_cs\_limit}, which sets the timestep based on the sound-crossing timescale 
for spatial zones near the shock front, between $0.1$ -- $10$. 
While \MESA{} is an implicit code and thus not limited by the sound-crossing timescale, it provides 
a physical and convenient timescale 
over which to probe the numerical sensitivity of our models.

As can be seen in Figure \ref{fig:mc_space}, increasing the mesh resolution by a factor of
5 shows changes smaller 
than those which result from the variations we tested in the injected energy or choice of mass cut.
Increasing (or decreasing) the temporal resolution also shows changes in ejected \carbon{} mass of $\lessapprox 0.05\%$ (Figure \ref{fig:mc_time}).  
Thus our numerical uncertainties during the supernova are much smaller than our
uncertainties due to either the stellar models or the physical explosion parameters.

These results suggest our \carbon{} estimates are therefore not 
sensitive to the uncertain parameters assumed for the explosion, however other isotopes which are formed 
deeper into the star are 
more affected by both the injected energy and the mass cut \citep{young07,suwa19,sawada19}. 
While our results show that the 
total amount of \carbon{} ejected is insensitive to the physical explosion parameters considered here, the amount 
of \carbon{} that is 
observable (via the production of carbon-rich dust at late times) is sensitive to the explosion 
physics \citep{brooker21}.

\section{IMF weighted yields}\label{sec:imf}

\begin{deluxetable*}{c|ccc}

  \tablehead{
    \colhead{BH formation assumption} & \multicolumn{3}{|c}{Ratio of \carbon{} yields (binary-stripped/single)} \\
   & \colhead{$\alpha=-1.9$} & \colhead{$\alpha=-2.3$} & \colhead{$\alpha=-2.7$}
  }
  
  \tablecaption{Ratio of the IMF weighted yields between an equal number of massive binary-stripped stars and single stars for different assumptions about the ejection of the envelope during core collapse and the IMF power-law $\alpha$, see also Equation \ref{eq:ratio}. In case of BH formation we assume that the carbon-rich layers fall back onto the BH.}
  \tablecolumns{4}
  \startdata
    All stars explode successfully & 1.51 & 1.44 & 1.36 \\
    BHs form from stars with initial masses $\rm{\mint>22}\msun$ & 1.79 & 1.55 & 1.35 \\
    BHs form from stars final core masses $\rm{M_{He,final}}>7\msun$ & 2.59 & 2.22 & 1.91 \\
   Schneider et al.\ 2021 like formation & 2.57 & 2.21 & 1.92 \\
  \enddata

\end{deluxetable*}\label{tab:imf_yields}

The limit between stars which form neutron stars (and are assumed to eject their envelopes) and those
that form black holes (which may or may not eject their envelopes) is uncertain \citep{oconnor:11,brown13}. Whether a star
ejects its envelope or not will strongly affect the final yields from that star. 
To probe this
uncertainty in whether a star ejects its envelope during core collapse we test several different ejection assumptions.
Table \ref{tab:imf_yields} show the IMF weighted ratio of the total \carbon{} yields for binary-stripped massive stars and single massive stars at solar metallicity.  We define the ratio as:

\begin{equation} \label{eq:ratio}
    \rm{Ratio = \frac{\int_{11}^{45} \left(Y_{b,ccsn}f_b + Y_{b,winds} + Y_{b,rlof} \right)
    M_{init,b}^{\alpha} \, dM}
    {\int_{11}^{45} \left(Y_{s,ccsn}f_s + Y_{s,winds} \right)
    M_{init,s}^{\alpha} \, dM}
    }
\end{equation}

Where we assumed a Salpeter-like IMF with different values of $\alpha$ \citep{schneider18}, $\rm{M_{init,b}}$ and $\rm{M_{init,s}}$ 
are the initial mass of the primary star in the 
binary and the mass of the single star in solar masses, $\rm{Y_b}$ and $\rm{Y_s}$ 
are the yields of \carbon{} in solar masses for the respective stars and for each type of mass loss, and
$\rm{f_b}$ and $\rm{f_s}$ are filter functions that 
are either 0 or 1 depending on whether
we assume that the star ejects its envelope. 
The integration limits are taken over the entire range of initial masses considered in this work.

Firstly assuming that all stars eject their envelopes at core-collapse then we find binary-stripped stars contribute $\approx40\%$ more \carbon{}
compared to the same initial mass of single massive stars, assuming a standard Salpeter IMF $\alpha=-2.3$. This is due to extra mass loss in binary systems, leading both to a higher \carbon{}
yield in the winds and in the final supernovae for $\mint\approx30 $ -- $ 40\msun$ 
binary-striped stars as compared to single stars.

As we do not expect 
all stars to eject their envelopes at core-collapse we can also filter out systems which we do not expect to eject their envelope.
\citet{sukhbold:14} found that single stars with $\mint \lessapprox 22\msun$ had a low compactness \citep{oconnor:11},
suggesting they would be likely to successfully eject their envelope, this would set the maximum helium core mass
as $\mhe \lessapprox 7\msun$. 

Assuming all stars eject their envelopes with $\mint < 22\msun$, then 
stripped primary stars in massive binaries contribute even more to
the \carbon{} in the Universe than single massive stars. 

\replaced{Finally taking}{Next, taking} the helium core mass at $\rm{M_{He,final}}<7\msun$, measured
at core collapse, binary-stripped massive stars eject approximately twice as much carbon than single massive stars (for $\alpha=-2.3$). 
This is due to the extra mass
loss binaries undergo, lowering the final helium core masses. 
Thus for a given helium core mass at core-collapse binary-stripped systems
have a higher initial mass \citep{kippenhahn:67,habets86}, expanding the range in initial masses over which we assume a successful explosion occurs
\citep{vartanyan21}. For our systems, a cut of $\rm{M_{He,final}}<7\msun$
is equivalent to a cut of $\mint<19\msun$ for single stars and $\mint<22\msun$ for binaries. 

Finally, we use the limits for BH formation found in Table 1 of \citet{schneider21}. They found that
NSs would form from two populations, firstly for single stars with $\mint\leq 21.5\msun$ and for stars which were stripped as case B binaries with
$\mint\leq 31.5\msun$. Then a second population of NSs are formed from more massive objects, for single stars in the range $23.5 \leq \mint/\msun \leq 34.0$ and for case B binaries with $34.0 \leq \mint/\msun \leq 67.5$. This second population arises from changes in the core carbon and neon burning, leading to differences in whether the burning is convective or radiative, which leads to differences in the final core compactness and structure \citep[see also][]{brown99,brown01}.

As the IMF $\alpha$ increases, thus favouring the production of more massive stars,
the contribution that binary-stripped makes to the \carbon{} yields increases. This is due
the greater weight now given to the \carbon{} yields from wind mass loss (as only the most massive stars in our grid contribute to wind mass loss yields), which is where the 
difference between the binary-stripped and single star yields is greatest (See Figure \ref{fig:yields}).

For all envelope assumptions the contribution from the wind mass loss is
greater for stripped binaries then for the single stars. This is due to both
the binary-stripped stars having \carbon{} positive yields at lower initial masses, and due to the higher \carbon{} yield at the equivalent initial masses
due to the extra mass loss from RLOF altering the core structure (Section \ref{sec:core}). Table \ref{tab:imf_yields2} in Appendix \ref{sec:table} shows how the \carbon{} yields from only core collapse vary as a function of the envelope ejection assumptions.
Adopting the prescription of \citet{schneider21} does not lead to a significantly different 
ratio of carbon yields from when using the simple helium core mass cut. Firstly, this is due to the ratio of the yields being dominated by the wind yield rather than the core-collapse yield, and secondly because increasing the initial mass range for successful envelope ejections increases both binary-stripped and single star core-collapse yields by approximately the same relative amount.

The actual contribution to carbon enrichment in the Universe also depends on an additional scaling factor weighting the fraction of stars that are single against those that are binary-stripped \citep{sana:12}, as well as 
the fraction of binary systems which do not self-strip. 
Note also that we are comparing equivalent total initial masses for ensembles of massive single stars and the one star in each binary that we model as stripped, not the full initial stellar mass of the binaries.
We have also not included the \carbon{} yields from the secondary stars in the binaries. 
If the secondary gains significant mass by accretion, we anticipate that this would further increase the relative efficiency of massive binary systems in producing \carbon{}.      

We can combine the result shown in Figure \ref{fig:yields}(b), indicating that the \carbon{} yields from winds of massive stars are not positive until $\mint\approx35\msun$, with the expectation that only massive stars with $\mint \lessapprox 22 \msun$ eject their envelopes at core collapse. This combination suggests that stars with initial masses $22\lessapprox \mint/ \msun \lessapprox 35$ do not
contribute to the net production of \carbon{} in the Universe.   The contribution of massive stars with $\mint < 22\msun$ can be well described by only their \carbon{} supernovae yield,
while stars with $\mint > 35\msun$ are well described by only their wind yields.

\section{Discussion}\label{sec:discus}

This work has only modelled solar-metallicity stars, has neglected the potential importance of rotational mixing, and our binary-stripped models all underwent early case B mass transfer. We have also tested only one set of physics assumptions for stellar evolution, even though there are many uncertainties in the evolution of massive stars. Here we discuss potential limitations from those approximations.

Wind mass loss prescriptions for massive stars are uncertain \citep{renzo:17}.
\citet{sander20b} found that we would expect weaker Wolf-Rayet winds, at solar 
metallicity, for stars in the $\rm{M=}10$ -- $15\msun$ range than for our choice of \citet{nugis:00}.
Lowering the wind mass-loss rate in this mass range would likely lead to lower supernova \carbon{} yields from binaries \citep{langer95,eldrigde04}.
As the stripped star would lose less mass during core helium burning, the helium core would recede less, leaving behind less \carbon{} in the helium shell.
This would decrease the significance of binary-stripped stars on the total \carbon{} yields. Single stars would be
unaffected as they do not self-strip until much higher masses, where the models of 
\citet{sander20b} agree with those of \citet{nugis:00} (until the initial masses go above those considered here,
where \citet{sander20b} would predict higher mass loss rates). See also \citet{dray03a} for
discussion of the effects of the choice of wind mass-loss prescription on carbon yields from Wolf-Rayet stars.

Correlated with the wind mass loss is the modelled metallicity.
Lower metallicity models would show a similar effect as having a lower assumed wind mass loss rate by changing the fraction of stars that become stripped \citep{eldridge06}. Also at low metallicities, RLOF in binaries may not fully strip the star due to changes in the envelope physics \citep{gotberg17,yoon:17,laplace20}. Without the envelope becoming fully stripped we would not see the helium core recede which would lower both the wind and core-collapse \carbon{} yields.

Our binary star models were each  given an initial period such that they would undergo early case B mass 
transfer, before the onset of core helium burning. However, binaries may also interact earlier during the main sequence (case A), 
or after core helium burning has started (case C). 
Mass lost during case A will primarily lead to the star having a smaller core mass \citep{wellstein99,yoon:10}.
If the mass loss ceases before the star undergoes core helium burning, and the star retains its hydrogen envelope,
then the helium core will not recede. As the core did not recede it will not leave behind
a pocket of \carbon{} above it.
Thus the winds from the binary-stripped star will not be enhanced in \carbon{} yields relative to single stars. 
Countering this, smaller cores would
lead to more successful explosions, ejecting the envelope, which acts to increase the total \carbon{} yields.
More work is needed to understand which effect dominates.
Mass lost during case C will occur after core helium burning has started. By this time the core mass of the star
is set, thus we would be unlikely to see an enrichment of the helium shell with \carbon{} from the helium core.

Rotation plays a key role in the evolution of stars \citep{meynet00} and the observed chemical composition of massive stars
\citep{hunter07,hunter08,hunter09}. 
The extra chemical mixing rotation 
generates inside the star can lead to larger core masses \citep{ekstrom12,murphy21}, as well as mixing 
elements to the surface of the star \citep{meynet06,maeder14,groh19}. \citet{prantzos18} found enhancements in \carbon{} of a 
approximately a factor of three, between initial rotation rates of $0$ -- $ 300 \kms$ for a $20\msun$ star.
\citet{hirschi04,hirschi05} found that the \carbon{} yields increase by a factor $1.5$ -- $2.5$ for stars 
$\mint<30\msun$, and with initial rotation rates of $0$ -- $ 300 \kms$ due to the more massive core that the rotating models produce. 

\citet{fields18} explore the sensitivity of a 15\msun{} solar metallicity star
to variations when all reaction rates in the model were varied within their measured
uncertainties. They found that at core helium depletion the central carbon fraction can
vary $\pm80\%$ compared to the model which adopts median reaction rates, driven
by variations in \crate{} and the triple-$\alpha$ rates. \citet{sukhbold20} showed
how the type of carbon burning, radiative or convective, is altered by changing
the \crate{} rate, in 14 -- 26\msun{} stars. The location of carbon ignition
plays a key role in determining how compact the core will become and thus how likely it is to successfully
explode \citep{brown99,brown01}, and thus whether it will
eject its envelope \citep{weaver93b,timmes96}. \citet{sukhbold20} found that variations in the
$\approx 1\sigma$ uncertainty in the \crate{} can move the
location between convective and non-convection carbon ignition by $\pm2\msun$
in initial mass. That change in initial masses is similar to the change in initial masses 
seen in Section \ref{sec:imf} when 
assuming stars with a fixed final helium core mass eject their envelopes. 

\section{Conclusion}\label{sec:conc}

Motivated by the high binary fraction inferred for young massive stars and the importance in improving chemical yield predictions, we have started a systematic investigation of the impact of binarity on the chemical yields of massive stars. In this paper, intended as the first in a series, we focus on \carbon{} yields. We present a systematic comparison of the differences between massive stars stripped by a binary companion and single stars with the same initial mass at solar metallicity.  

To achieve this, we modelled the evolution of from the onset of hydrogen burning until core collapse. We then followed the nucleosynthesis through the resulting supernova shock and compute the ejected \carbon{} yield for different mass loss processes. 

Our results can be summarised as follows:

\begin{enumerate}

\item We find that massive stars stripped in binaries during hydrogen-shell burning are nearly twice as efficient at contributing to the cosmic \carbon{} production as a similar number of massive single stars (a factor 1.4--2.6 depending on the assumptions for black hole formation and the slope of the initial mass function, see Equation ~\ref{eq:ratio} and Table~\ref{tab:imf_yields}). 

\item We confirm that the difference in yields between binary-stripped and single massive stars can be explained by considering the outer-most \carbon{}-rich layers produced in the early phases of central helium burning. In single stars these layers tend to get mixed into the growing convective core, leading to destruction of \carbon{} by alpha captures and later carbon burning. In binary-stripped stars (and in single stars stripped by stellar winds) the convective helium-burning core cannot grow and may even retreat. The outer-most \carbon{}-rich layers disconnect and form a pocket where the temperature never becomes high enough for \carbon{} destruction (Figure ~\ref{fig:kipps}), not even when a supernova shock wave passes through and ejects these layers (Figure ~\ref{fig:shock}) \cite[cf.][]{langer91, woosley19, laplace21}.

\item Stellar winds also eject \carbon{} once the carbon-rich layers are exposed to the surface. This only happens for the most massive stars in our grid (initial masses $>36\Msun$ and $>38\Msun$ for our solar-metallicity binary-stripped and single stars respectively). Comparing stars of the same initial mass, we find that the wind yields for binary-stripped stars are higher because the carbon-rich layers are more massive and appear earlier at the surface (see Figure~\ref{fig:yields} and \ref{fig:yields}b).

\item Mass loss from binary systems during conservative mass transfer does not contribute to \carbon{} yields, in our models. These layers are either pristine or contain CNO-processed material which is \carbon{} poor \citep[c.f.][]{demink09a}.

\item Our yield predictions are remarkably robust with respect to choices in the treatment of the explosion, such as the explosion energy, how the energy is inserted, and the mass that promptly forms a compact object. These lead to variations smaller than $\approx0.5\%$ (See Figure ~\ref{fig:mc_sn}).

\item We show that the \carbon{} yield predictions of both single and binary-stripped stars are sensitive to the treatment of overshooting, specifically during carbon-shell burning.  We find variations up to $\approx40\%$ in the final carbon yields for individual models (see Figure \ref{fig:coreo_c12}), depending on the size and lifetime of the shells.  We identify this as a primary source of uncertainty and cause for noise in our predictions of the supernova yields. 

\item We note that the variations between the predictions of \carbon{} yields for massive stars presented in other recent studies \citep[][]{pignatari16,limongi18, griffith21} is large and of a similar order as the differences we find between binary-stripped and single stars. 

\end{enumerate}

We conclude that the yields of massive binary-stripped stars are systematically different from those of massive single stars. The effects of binarity should therefore not be ignored if we want to obtain more reliable yield predictions for carbon and better understand the relative contribution of massive stars with respect to the other proposed sources of carbon, namely Asymptotic Giant Branch stars and type Ia supernova. 

However, the main priority would be to better understand how to treat boundary mixing, especially during the carbon shell burning phase, as this seems to be the primary uncertainty in both single and binary star models. Further aspects, which deserve further investigation, include the effects of metallicity and how these interplay with uncertainties in the stellar wind mass loss, and the still unknown conditions for which stars lead to successful supernova explosions. Advancing in these areas would require efforts on both the theory and observation side.

Finally, we emphasize that the binary-stripped stars studied here represent only one of the many possible final fates for a star born in the close vicinity of a companion. Binary interaction can affect massive stars in many other ways. This study is thus only a first step towards answering the bigger question of how binarity impacts the chemical yields of stars.

\appendix

\section{Other physics choices}\label{sec:other_phys}

The EOS in \MESA{} is blended from several sources; OPAL \citep{Rogers2002}, SCVH
\citep{Saumon1995}, PTEH \citep{Pols1995}, HELM
\citep{Timmes2000}, and PC \citep{Potekhin2010}.
Opacities are primarily from OPAL \citep{Iglesias1993,
Iglesias1996} with additional data from \citet{Buchler1976,Ferguson2005,Cassisi2007}

Nuclear reaction rates are a combination of rates from
NACRE  and JINA's REACLIB \citep{Angulo1999,Cyburt2010}, with
additional weak reaction rates from \citet{Fuller1985, Oda1994,
Langanke2000}. Nuclear screening
is computed with the prescription of \citet{Chugunov2007}.  Thermal
neutrino loss rates are from \citet{Itoh1996}. We compute the
Roche lobe radii in binary systems using the fit from \citet{Eggleton1983}.
The mass transfer rate in our Roche lobe overflowing binary systems is computed from  
the prescription of \citet{Ritter1988}.

\section{Comparison to other works}

\subsection{Laplace et al.\ 2021}\label{sec:comp}

\begin{figure}[]
  \centering
  \includegraphics[width=1.0\linewidth]{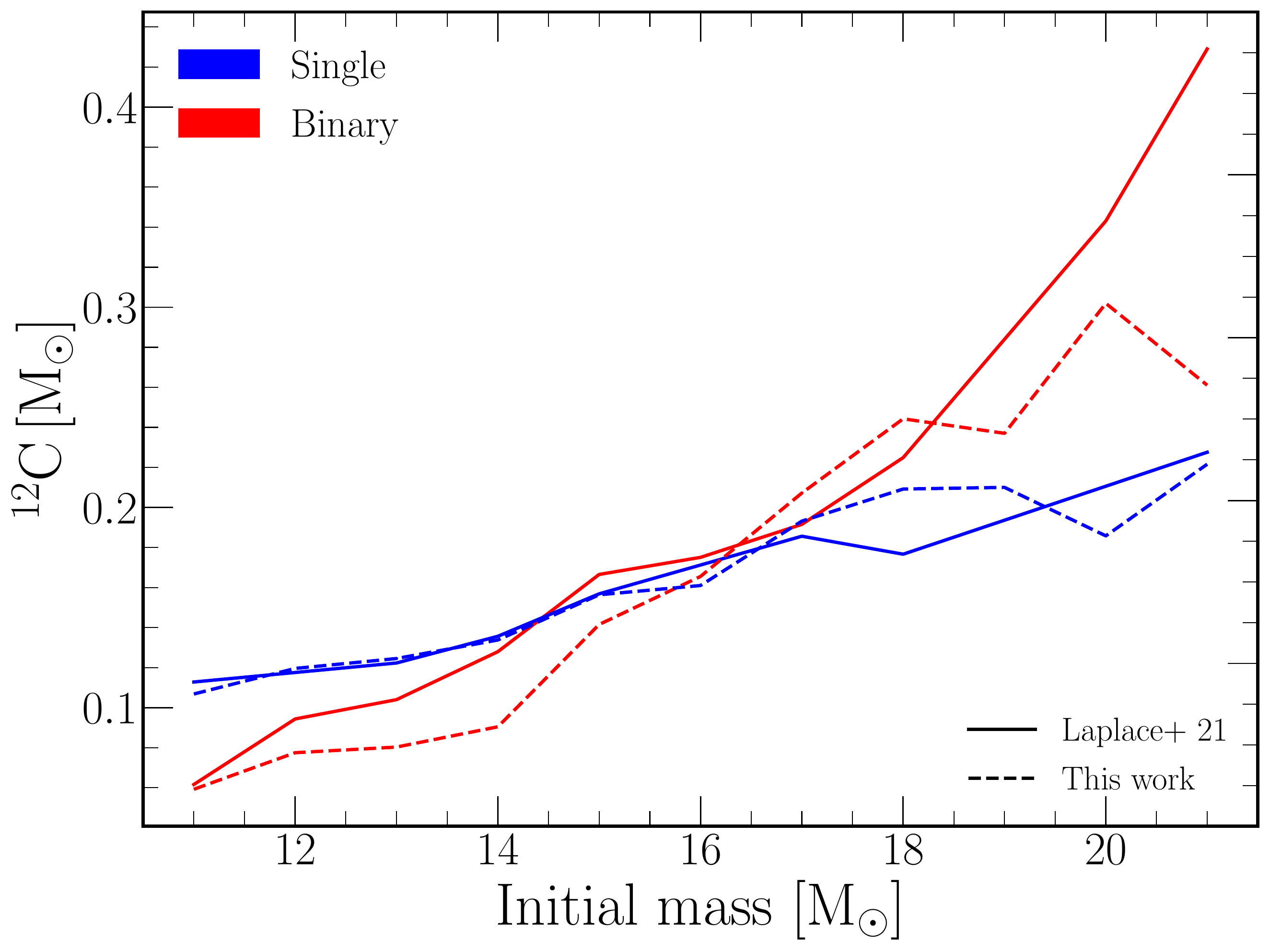}
  \caption{Total \carbon{} mass ejected during supernovae, comparing models from this work (dashed) with those from \citet{laplace21} (solid). In blue are plotted single-star models, while red represents binary-stripped models. 
  The models of \citet{laplace21} were evolved with \texttt{mesa\_128.net} after core oxygen depletion, and then exploded with the same method as in Section \ref{sec:meth}.}
  \label{fig:eva}
\end{figure}

In Figure \ref{fig:eva} we compare the \carbon{} mass ejected in the supernova calculations presented in Figure \ref{fig:yields} to the \carbon{} ejected when using progenitor models computed in \citet{laplace21}.
We took the models from \citet{laplace21} and exploded them with the same method as in 
Section \ref{sec:meth}. There are two primary differences between these sets of models;
\citet{laplace21} was computed with \MESA{} r10398, while this work uses \MESA{} r11215; and
\citet{laplace21} uses \texttt{mesa\_128.net} after core oxygen depletion, while this work continues to use \texttt{approx21.net} after core oxygen depletion. For both sets of models we explode them with \MESA{} r11215.

We can see that for both single and binary-stripped star models the amount of \carbon{} ejected in 
the supernovae is similar between both sets.  The largest disagreement occurs for the binary-stripped stars at $\mint=21\msun$. This is due to the carbon shell destroying more \carbon{} in the pre-supernova phase
of the evolution. As this carbon shell acts before the switch to the larger nuclear network this indicates the differences are due to changes in \MESA{} between r10398 and r11215, rather than to differences in the nuclear network.

This indicates
that it is reasonable, when considering the \carbon{} yields, to use a smaller but more computational efficient, nuclear network during the pre-supernova evolution. This is due to two reasons, first most of the \carbon{} that will be ejected is set by the evolution up to the end of core-carbon burning, which does not require large nuclear networks to compute. Secondly, the \carbon{} that is ejected in the supernova resides in/near the helium shell. Thus the total \carbon{} yield is insensitive to the 
structure of the inner core, which is sensitive to the choice of nuclear network \citep{farmer16}. This however may not be true for
other isotopes, or for inferring neutron star masses which depend sensitively on pre-supernova structures.

\subsection{Other core-collapse works}\label{sec:others}

\begin{figure}[h]
  \centering
  \includegraphics[width=1.0\linewidth]{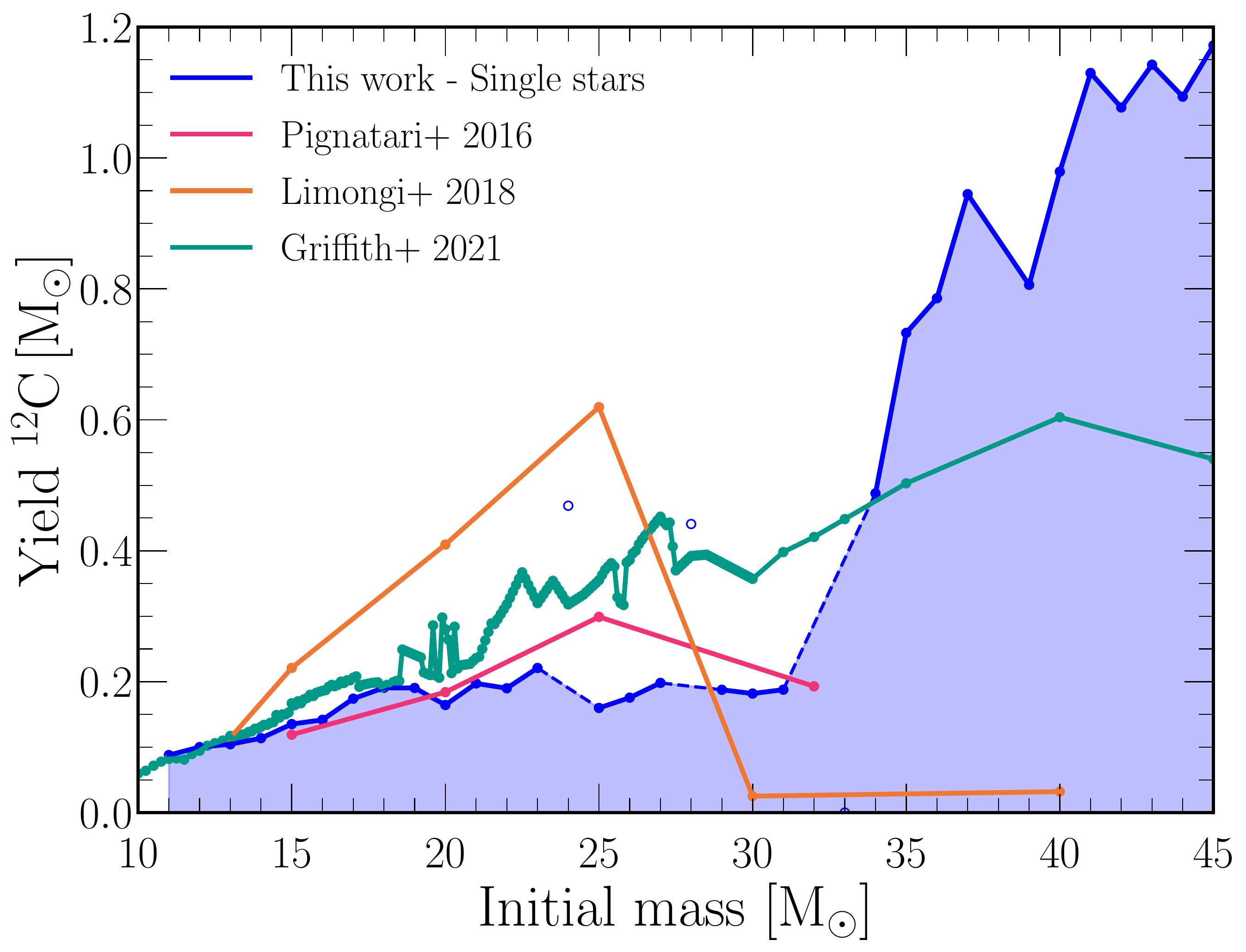}
  \caption{Comparison of nucleosynthetic \carbon{} yields for solar metallicity non-rotating single stars. 
  In blue this work, 
  magneta line \citep{pignatari16} (delayed model), orange line \citep{limongi18}, and teal line \citep{griffith21}. Dashed
  lines and empty circles have the same meaning as in Figure \ref{fig:yields}}
  \label{fig:comp}
\end{figure}
 
Figure \ref{fig:comp} shows a comparison between our core collapse yields and the 
results of 
\citet{pignatari16,limongi18} and \cite{griffith21} for solar metallicity non-rotating single stars. 
Below $\mint \approx20\msun$ there is a reasonable agreement between the different models, 
and software instruments, for the final core-collapse \carbon{} yield. Above this point the different models
begin to diverge, most likely due to changes in the behaviour of carbon shells in each
set of models and choice of nuclear reaction rates. Differences in the choices for the treatment of stellar wind mass loss likely also plays a role. 
We note that the spread in \carbon{} yields is of the order of the spread we find due to the differences
in carbon shell burning (see section \ref{sec:presn_sen}). Note the two open symbols, which indicate models in our grid where the mixing due to convective carbon shell burning destroyed less carbon. These two models are more consistent with the higher yields found by \citet{limongi18} and \citet{griffith21}. 

\section{Table of yields}\label{sec:table}

In Table~\ref{tab:yields} we show the total \carbon{} yields based on the source of the mass loss. 
Several of the models show anomalous carbon burning (see section \ref{sec:cshell}), however
the single $\mint=33\msun$ star shows a slightly different behaviour. Here the convection zone above the carbon-burning shell extended sufficiently far 
that the mixing region dredged down all the \carbon{} that existed in/near the helium shell. Thus by the time of core-collapse 
the star becomes depleted in \carbon{}.

We also provide the integrated yields for single and binary-stripped stars  for different assumption with respect which stars explode successfully in Table~\ref{tab:imf_yields2}.  The results are normalized to the assumption that all stars explode successfully. We can see that changing the assumption for the successful explosion has a large impact on the final \carbon{} yields, with the yields decreasing by ~60--70\%. The binary-stripped models show no differences between the different envelope assumptions as the final helium core mass cut and cut on initial masses are equivalent.

\begin{deluxetable*}{c|cccc|ccccc}
\tablewidth{\linewidth}
\tablecaption{\carbon{} yields in solar masses broken down by mass-loss type for both single stars and the primary star in 
the binary. $\rm{M_{He,final}}$ is the final helium core mass at core-collapse . \label{tab:yields} }

\tablehead{
\colhead{Initial mass [\msun]} & \multicolumn{4}{c}{Single [\msun]} & \multicolumn{5}{c}{Binary [\msun]} \\
  & $\rm{M_{He,final}}$ & $\rm{M_{Winds}} $ & $\rm{M_{CC}}$ & $\rm{M_{Total}}$ & $\rm{M_{He,final}}$ & $\rm{M_{RLOF}}$ & $\rm{M_{Winds}} $ & $\rm{M_{CC}} $ & $\rm{M_{Total}} $ \\  
}

\startdata
11 &    3.798 &   -0.001 &    0.088 &    0.087 &    3.191 &   -0.006 &   -0.001 &    0.055 &    0.051 \\
12 &    4.297 &   -0.002 &    0.100 &    0.098 &    3.602 &   -0.007 &   -0.001 &    0.073 &    0.069 \\
13 &    4.800 &   -0.002 &    0.104 &    0.102 &    4.014 &   -0.007 &   -0.002 &    0.074 &    0.071 \\
14 &    5.308 &   -0.003 &    0.114 &    0.111 &    4.422 &   -0.007 &   -0.002 &    0.084 &    0.081 \\
15 &    5.829 &   -0.004 &    0.135 &    0.132 &    4.832 &   -0.007 &   -0.002 &    0.134 &    0.132 \\
16 &    6.359 &   -0.004 &    0.142 &    0.137 &    5.222 &   -0.008 &   -0.003 &    0.157 &    0.156 \\
17 &    6.896 &   -0.005 &    0.174 &    0.169 &    5.608 &   -0.008 &   -0.003 &    0.198 &    0.198 \\
18 &    7.414 &   -0.006 &    0.191 &    0.185 &    5.980 &   -0.008 &   -0.004 &    0.235 &    0.235 \\
19 &    7.817 &   -0.007 &    0.190 &    0.184 &    6.243 &   -0.008 &   -0.004 &    0.226 &    0.227 \\
20 &    8.275 &   -0.007 &    0.164 &    0.157 &    6.579 &   -0.009 &   -0.005 &    0.290 &    0.291 \\
21 &    8.382 &   -0.007 &    0.197 &    0.190 &    6.793 &   -0.010 &   -0.005 &    0.249 &    0.251 \\
22 &    7.943 &   -0.009 &    0.190 &    0.181 &    8.365 &   -0.011 &   -0.002 &    0.134 &    0.133 \\
23 &    9.366 &   -0.009 &    0.221 &    0.212 &    8.246 &   -0.011 &   -0.004 &    0.233 &    0.234 \\
24 &    9.856 &   -0.010 &    0.469\tablenotemark{a} &    0.458\tablenotemark{a} &    9.476 &   -0.008 &   -0.004 &    0.178 &    0.182 \\
25 &    9.530 &   -0.011 &    0.160 &    0.149 &    9.403 &   -0.010 &   -0.004 &    0.250 &    0.256 \\
26 &    9.995 &   -0.011 &    0.176 &    0.164 &   10.636 &   -0.010 &   -0.005 &    0.185 &    0.191 \\
27 &   10.785 &   -0.013 &    0.198 &    0.185 &   10.283 &   -0.009 &   -0.005 &    0.187 &    0.196 \\
28 &   11.154 &   -0.012 &    0.441\tablenotemark{a} &    0.429\tablenotemark{a} &   10.889 &   -0.009 &   -0.006 &    0.189 &    0.201 \\
29 &   12.194 &   -0.014 &    0.188 &    0.173 &   11.085 &   -0.009 &   -0.010 &    0.613 &    0.631 \\
30 &   12.449 &   -0.014 &    0.182 &    0.167 &   11.094 &   -0.010 &   -0.011 &    0.719 &    0.737 \\
31 &   13.303 &   -0.015 &    0.188 &    0.173 &   13.131 &   -0.011 &   -0.007 &    0.167\tablenotemark{a} &    0.182\tablenotemark{a} \\
32 &   14.238 &   -0.017 &    0.000\tablenotemark{b} &    0.000\tablenotemark{b} &   11.632 &   -0.010 &    0.012 &    0.949 &    0.997 \\
33 &   15.031 &   -0.016 &    0.000\tablenotemark{a} &   -0.016\tablenotemark{a} &   11.969 &   -0.011 &   -0.013 &    0.866 &    0.890 \\
34 &   14.870 &   -0.019 &    0.488 &    0.469 &   12.818 &   -0.010 &   -0.014 &    0.718 &    0.747 \\
35 &   15.031 &   -0.021 &    0.733 &    0.712 &   13.252 &   -0.010 &    0.025 &    1.002 &    1.073 \\
36 &   15.469 &   -0.021 &    0.786 &    0.764 &   11.650 &   -0.014 &    0.575 &    0.975 &    1.595 \\
37 &   15.789 &   -0.005 &    0.945 &    0.939 &   13.159 &   -0.010 &    0.879 &    0.977 &    1.912 \\
38 &   16.196 &    0.052 &    0.000\tablenotemark{b} &    0.000\tablenotemark{b} &   13.552 &   -0.010 &    0.962 &    1.008 &    2.029 \\
39 &   16.576 &    0.120 &    0.806 &    0.926 &   13.948 &   -0.011 &    1.039 &    0.971 &    2.071 \\
40 &   16.966 &    0.186 &    0.979 &    1.165 &   14.357 &   -0.011 &    1.112 &    1.063 &    2.240 \\
41 &   17.311 &    0.277 &    1.130 &    1.407 &   14.776 &   -0.011 &    1.180 &    1.062 &    2.311 \\
42 &   17.638 &    0.379 &    1.077 &    1.456 &   15.183 &   -0.012 &    1.253 &    1.021 &    2.346 \\
43 &   17.999 &    0.458 &    1.143 &    1.600 &   15.599 &   -0.013 &    1.322 &    1.134 &    2.531 \\
44 &   18.328 &    0.554 &    1.094 &    1.648 &   15.831 &   -0.013 &    1.434 &    1.069 &    2.585 \\
45 &   18.621 &    0.667 &    1.172 &    1.839 &   16.279 &   -0.013 &    1.485 &    1.138 &    2.709 \\
\enddata
\tablenotetext{a}{Model shows anomalous carbon burning behaviour (See section \ref{sec:cshell})}
\tablenotetext{b}{Model did not reach core-collapse.}
\end{deluxetable*}

\begin{deluxetable*}{c|cc}

  \tablehead{
    \colhead{BH formation assumption} & \multicolumn{2}{|c}{Normalised \carbon{} yields } \\
   & \colhead{Single stars} & \colhead{Binary-stripped stars} 
  }
  
  \tablecaption{ Impact of assumptions regarding black hole formation on the core-collapse IMF weighted yield predictions normalised to the default assumption that all stars explode successfully.  We assume an IMF power-law exponent $\alpha=-2.3$. In case of BH formation we assume that the carbon-rich layers fall back onto the BH.}
  \tablecolumns{3}
  \startdata
    All stars explode successfully & 1.0 & 1.0  \\
    BHs form from stars with initial masses $\rm{\mint>22}\msun$ & 0.41 & 0.34  \\
    BHs form from stars final core masses $\rm{M_{He,final}}>7\msun$ & 0.27 & 0.34  \\
    Schneider et al 2021 like formation & 0.55 & 0.84 \\
  \enddata

\end{deluxetable*}\label{tab:imf_yields2}

\begin{acknowledgments}
We acknowledge helpful discussions with B.~Paxton, F.~Timmes,
L.~van Son, and T.~Wagg. 
The authors acknowledge funding by the European Union's Horizon 2020 research
and innovation program from the European Research Council (ERC)
(Grant agreement No.\ 715063), and by the Netherlands Organization for
Scientific Research (NWO) as part of the Vidi research program BinWaves
with project number 639.042.728 and a top module 2 grant with project number 614.001.501.
This work was also supported by the Cost Action Program ChETEC CA16117.
This work was carried out on the Dutch national e-infrastructure
with the support of SURF Cooperative. This research has made use of NASA's
Astrophysics Data System.
\end{acknowledgments}

 \software{
 \texttt{mesaPlot} \citep{mesaplot},
 \texttt{mesaSDK} \citep{mesasdk},
 \texttt{ipython/jupyter} \citep{perez_2007_aa,kluyver_2016_aa},
 \texttt{matplotlib} \citep{hunter_2007_aa},
 \texttt{NumPy} \citep{der_walt_2011_aa},
 \MESA \citep{paxton:11,paxton:13,paxton:15,paxton:18,paxton:19}, and 
 \texttt{pyMesa} \citep{pymesa}.
          }

\bibliographystyle{aasjournal}
\bibliography{ccsn}

\end{document}